\documentclass[journal, twoside]{IEEEtran}
\usepackage[table]{xcolor}
\usepackage{indentfirst}
\usepackage{graphicx}
\usepackage{amsmath}
\usepackage{amssymb}
\usepackage{amsfonts}
\usepackage{mathrsfs}
\usepackage{leftidx}
\usepackage{color}
\usepackage{amsmath}
\usepackage{arydshln}
\usepackage{amsthm}
\usepackage{ragged2e}
\usepackage{cite}
\usepackage{enumerate}
\usepackage{longtable}
\usepackage{float}
\usepackage{stfloats}
\usepackage{hyperref }
\usepackage{algpseudocode}
\usepackage{algorithm}
\usepackage[caption=false,font=normalsize]{subfig}
\usepackage{tabularx}
\usepackage{booktabs} % 在导言区加这个宏包
\usepackage{multirow} 

\theoremstyle{plain}

\newtheorem{rem}{Remark}

\newcommand{\RMnum}[1]{\uppercase\expandafter{\romannumeral #1}}

\usepackage{caption}

\newcolumntype{P}[1]{>{\raggedright\arraybackslash\footnotesize}m{#1}}
\newcolumntype{A}[1]{>{\centering\arraybackslash\footnotesize}m{#1}}

\usepackage[table,usenames,dvipsnames]{xcolor}
\definecolor{aa}{RGB}{175,238,238}
\definecolor{bb}{RGB}{255,255,255}
\usepackage{bm}
\usepackage{makecell}

\begin{document}

\title{APEG: Adaptive Physical Layer Authentication with Channel Extrapolation and Generative AI}

\author{Xiqi Cheng, Rui Meng,~\IEEEmembership{Member,~IEEE,} Xiaodong Xu,~\IEEEmembership{Senior Member,~IEEE,} Haixiao Gao, 

Ping Zhang,~\IEEEmembership{Fellow,~IEEE} and Dusit Niyato,~\IEEEmembership{Fellow,~IEEE}

\thanks{
This work was supported in part by the National Key Research and Development Program of China under Grant 2020YFB1806905; in part by the National Natural Science Foundation of China under Grant 62501066 and U24B20131; and in part by the Beijing Natural Science Foundation under Grant L242012. 
\textit{(Corresponding author: Rui Meng.)}

Xiqi Cheng, Rui Meng, Xiaodong Xu, Haixiao Gao, and Ping Zhang are with the State Key Laboratory of Networking and Switching Technology, Beijing University of Posts and Telecommunications, Beijing 100876, China (email: chengzi@bupt.edu.cn; buptmengrui@bupt.edu.cn; xuxiaodong@bupt.edu.cn; haixiao@bupt.edu.cn; pzhang@bupt.edu.cn).

Dusit Niyato is with College of Computing and Data Science, Nanyang Technological University, Singapore (email: dniyato@ntu.edu.sg).
}}

\maketitle

\begin{abstract}
With the rapid advancement of 6G, identity authentication has become increasingly critical for ensuring wireless security. The lightweight and keyless Physical Layer Authentication (PLA) is regarded as an instrumental security measure in addition to traditional cryptography-based authentication methods. However, existing PLA schemes often struggle to adapt to dynamic radio environments. To overcome this limitation, we propose the Adaptive PLA with Channel Extrapolation and Generative AI (APEG), designed to enhance authentication robustness in dynamic scenarios. Leveraging Generative AI (GAI), the framework adaptively generates Channel State Information (CSI) fingerprints, thereby improving the precision of identity verification.
To refine CSI fingerprint generation, we propose the Collaborator-Cleaned Masked Denoising Diffusion Probabilistic Model (CCMDM), which incorporates collaborator-provided fingerprints as conditional inputs for channel extrapolation. Additionally, we develop the Cross-Attention Denoising Diffusion Probabilistic Model (CADM), employing a cross-attention mechanism to align multi-scale channel fingerprint features, further enhancing generation accuracy.
Simulation results demonstrate the superiority of the APEG framework over existing time-sequence-based PLA schemes in authentication performance. Notably, CCMDM exhibits a significant advantage in convergence speed, while CADM, compared with model-free, time-series, and VAE-based methods, achieves superior accuracy in CSI fingerprint generation. The code is available at https://github.com/xiqicheng192-del/APEG

\end{abstract}


\section{Introduction}
%polish the english language "XX"
With the rapid development of the 6th generation (6G) communication technology, it will enable the blueprint of connecting everything, full-dimensional coverage, technology cross-integration, and human-like intelligent autonomy \cite{r1}. However, more open and heterogeneous wireless environments will introduce unprecedented new security challenges \cite{r2}. Consequently, as an important line of defense for wireless security, identity authentication becomes especially critical in 6G. Traditional identity authentication relies on cryptographic algorithms, such as 5G AKA \cite{r5}. However, it will face many limitations in future 6G, with the feature of high mobility, wide coverage, and low-power devices. Specifically, cryptography-based authentication schemes require key generation, distribution, updating, and revocation, making the whole process complex and computational intensive. Moreover, key lifetimes are short, and the updates are frequent, which raises the management burden and makes it difficult to meet strict latency targets. Furthermore, low-power devices lack the processing power for strong cryptographic operations \cite{r7,meng2025survey}.

As significant complement for traditional cryptography-based authentication techniques, Physical Layer Authentication (PLA) achieves reliable identity recognition by leveraging physical layer features, such as Channel State Information (CSI) fingerprints \cite{zhang2025physical,meng2023physical, duan2025adaptive} and Radio Frequency (RF) fingerprints \cite{r9,r10}. Since PLA avoids computationally-intensive encryption and frequent key updates, it is suitable for 6G scenarios filled with many low power devices.
% As one of PLA methods, Radio frequency (RF) fingerprints come from tiny differences in the RF front-end circuits caused by hardware manufacturing. These differences create unique characteristics of the device, such as carrier frequency offset (CFO) [13] and I/Q imbalance (IQI) [14][10]. However, since RF fingerprints are assumed to be stable, they are vulnerable to replay attacks [11]. Moreover, in dynamic scenarios, signal changes caused by multipath, blockage, or movement may also make RF fingerprints unstable. Moreover, these features rely heavily on static hardware traits. As a result, under low Signal-to-Noise Ratio (SNR) or strong interference, differences between devices may become hard to detect. Unlike RF fingerprints,
As crucial identity characteristics for PLA, CSI fingerprints are determined by the location and surrounding environment of the devices. Due to the inherent decorrelation features of CSI fingerprints, even a location change of half a wavelength can cause clear differences in received signal patterns \cite{r11}. This means devices at different physical positions will observe distinct CSI fingerprints, which are hard to fake or mimic. Moreover, CSI fingerprints are highly sensitive to both time and environments, making it difficult to reproduce the same pattern at different times or locations. This inherent variability provides strong resistance to replay attacks \cite{r12}. Therefore, as environment-perceptive characteristics, CSI fingerprints are suited to build location-aware authentication \cite{r13, r14, r15}.

Early studies on CSI-based PLA mainly focused on static scenarios. One common approach was to combine statistical channel models with hypothesis testing to perform identity verification \cite{r16,r17, ji2025physical}. These methods typically analyzed the statistical properties of CSI fingerprints and determined decision thresholds for PLA. With the rise of Machine Learning (ML), it was employed to improve threshold selection and adaptability \cite{r18}. Additionally, some researchers adopted a data-driven view and analyzed CSI features directly for classification. Support Vector Machines (SVM) \cite{r19,r20} and Gaussian Mixture Models (GMM) \cite{r21,r22} were employed as classification algorithms to distinguish devices by learning discriminative patterns from their CSI fingerprints. Furthermore, Deep Learning (DL) was introduced to capture spatial and temporal features of CSI fingerprints \cite{r23,r24}. ML demonstrates strong generalization over diverse channel conditions and effective authentication performance without requiring explicit modeling of channel distributions \cite{lai2025comprehensive,meng2025survey}.

Although significant progress has been made in adapting PLA to dynamic wireless environments \cite{r25, r26,r35,R36}, achieving robust and continuous authentication under rapidly time-varying channel conditions remains a critical challenge. In such environments, fast channel dynamics make it difficult for hypothesis testing methods based on statistical models to maintain reliable performance with fixed detection thresholds, resulting in degraded authentication accuracy. Moreover, traditional ML methods based on classifiers assume a stable mapping between CSI fingerprints and spatial locations of devices. However, frequent device movement or rapid environmental changes can cause significant shifts in the distribution and temporal correlation of CSI fingerprints, making the learned decision boundaries difficult to generalize and ultimately degrading authentication performance \cite{meng2023multiuser,meng2023physical}. Although Pan et al. \cite{r25} attempted to model time-varying CSI features using ML and applied a binary classification approach to avoid threshold-setting issues in dynamic scenarios, this method remained unstable in practical use. It was also highly sensitive to the number and spatial distribution of training fingerprint samples, which limited its generalization capability under dynamic conditions. Meanwhile, time-series DL methods have been proposed to address channel variations in dynamic environments \cite{r26,r35,R36}. However, CSI fingerprints are high-dimensional and structurally complex data, and they are highly susceptible to channel estimation errors, particularly under low SNR conditions. Consequently, existing time-series models still face difficulties in accurately tracking dynamic channels or achieving reliable authentication in dynamic environments.

To address the challenges mentioned above, we propose a PLA framework that combines channel extrapolation and Generative AI (GAI). 
Channel extrapolation explores correlations of wireless channels across time, frequency, space, and user domains. 
It extends the limited CSI collected by pilot signals to other time points, frequency bands, antennas, or spatial locations. This technique allows high-resolution channel estimation without increasing pilot overhead, which supports real-time sensing of dynamic wireless conditions\cite{r27,r28}. 
Earlier studies on channel extrapolation mainly relied on geometric parametric models and sparse reconstruction algorithms \cite{r29,r30}. More recently, DL models have shown strong potential to capture complex intra- and cross-domain CSI mappings. These models provide higher prediction accuracy and stronger robustness against time-varying channel conditions \cite{r31,r32}.
In this paper, we leverage spatial correlations among cooperative devices to design a prior inference mechanism based on channel extrapolation. With this mechanism, we can infer the expected CSI of the target device (Alice) using the known CSI of nearby collaborative device (Jack), which enables further authentication decisions.
In addition, we apply GAI to more accurately capture the structure of complex wireless channels and to better model the relationships between neighboring channels. Compared with Discriminative AI (DAI), GAI, such as GANs, VAEs, and Denoising Diffusion Probabilistic Models (DMs), learns the underlying data distribution and builds a mapping between the latent space and the observation space \cite{cao2024survey}. It can, therefore, generate diverse and high-quality samples, even when data is sparse or partially missing \cite{r33}.
GAI has recently emerged as a promising technology for enhancing Physical Layer Security (PLS), specifically in authentication tasks. For instance, Generative Adversarial Networks (GANs) have been successfully employed to authenticate RF transmitters by utilizing a generator to produce synthetic signals and a discriminator to distinguish between real and fake signals, achieving a detection accuracy of 99\%, which significantly outperforms traditional CNN-based approaches  \cite{roy2019detection}. Similarly, Variational Autoencoders (VAEs) have been applied to extract robust latent features from high-dimensional channel impulse responses for device authentication in Industrial IoT scenarios, effectively addressing the lack of prior information about attackers \cite{meng2022physical}. Furthermore, GAI has been utilized to augment training datasets with adversarial samples, thereby bolstering the classifier's robustness against spoofing and jamming attacks  \cite{merchant2019securing}.
Overall, the proposed approach can significantly improve identification performance in dynamic environments by combining channel extrapolation for spatial inference and GAI for structure-aware CSI generation.

% In recent years, frameworks such as GANs, VAEs, and diffusion models have been widely used in areas like channel feedback [33], image inpainting [34], and task-oriented communication [35]. 
% These models show strong ability to represent complex distributions and maintain robustness under dynamic conditions.
% In this paper, we develop a novel PLA framework based on channel fingerprints aiming to address the identity authentication  challenges in mobile 6G scenarios. We design an authentication architecture that combines channel extrapolation with generative modeling. This enables accurate channel prediction and efficient authentication for mobile devices such as Alice. 
% This design aims to improve the accuracy and robustness of device identification in dynamic environments. 
The main contributions are summarized as follows.
\begin{enumerate}
    \item We propose an Adaptive PLA Scheme with Channel Extrapolation and GAI (APEG) to realize reliable identity authentication in dynamic environments. By leveraging CSI fingerprints of nearby collaborative devices as the condition of channel extrapolation, the GAI model can adaptively generate real-time CSI fingerprints of legitimate users to improve the accuracy of their identity characteristics.
    \item Upon the proposed APEG approach, we transform CSI fingerprints into images and employ the Collaborator-Cleaned Masked Denoising Diffusion Probabilistic Model (CCMDM) for fingerprint generation. This model injects Gaussian noises into Alice's fingerprints as mask during the forward process, and reconstructs fingerprints in the reverse denoising process by leveraging collaborator's fingerprints. This approach benefits from low training complexity, making it suitable for practical deployment.
    \item We further propose the Cross-Attention Denoising Diffusion Probabilistic Model (CADM) to enhance the generation accuracy of fingerprints. Compared with the proposed CCMDM-based fingerprint generation scheme, the proposed CADM-based scheme employs cross-attention mechanism to fuse multi-scale features from Alice's and Jack's fingerprints, resulting in more accurate CSI fingerprint generation.
    \item We conduct simulations on the publicly available DeepMIMO datasets \cite{Alkhateeb2019}. The simulation results demonstrate the superiority of the proposed APEG method in authentication performance than Long Short-Term Memory (LSTM)-based and Gated Recurrent Unit (GRU)-based schemes. Additionally, the proposed CADM-based fingerprint generation scheme outperforms the scheme using Collaborator's fingerprints as Alice's fingerprints (CA) and the Cross-Attention VAE (CAVAE)-based scheme. Although the proposed CMDM-based scheme yields slightly lower fingerprint generation accuracy, it significantly increases convergence speed compared to the CADM-based scheme in training stage.
\end{enumerate}

% The remainder of the paper is organized as follows. In Section \RMnum{2}, we introduce the PLA model based on the channel of the neighbouring device Jack. Two proposed Gnerative models, which are used to learn the mapping relationship between the CSI of Alice and that of Jack, are explained in Section \RMnum{3}. Moreover, The simulation results are presented in Section \RMnum{4}. Finally, we conclude this paper in Section \RMnum{5}

% 1、提出了一个基于信道外推特性和生成式模型的PLA框架，用于动态场景（SNR在变、位置在变）、不需要攻击者的先验数据集。2、基于mask diffusion的指纹生成模型。3、基于attention-diffusion的指纹生成模型。4、仿真

% \section{Related Works}

\section{System Model and Problem Formulation}

\subsection{Network Model}
\begin{figure}[t]
    \centering
    \includegraphics[width=0.5\textwidth,trim=0 0 0 0,clip]{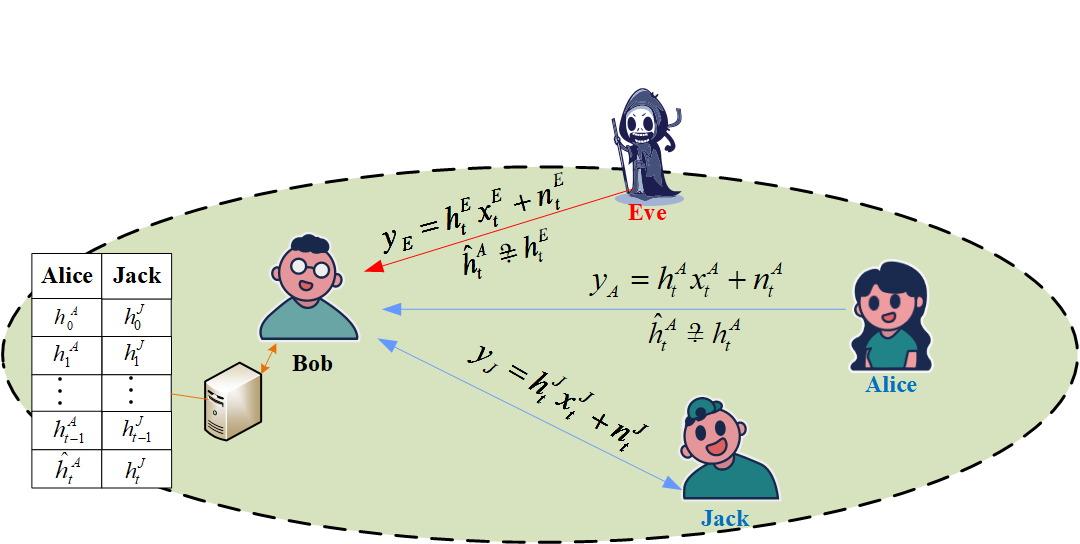}
    \caption{The system model, where the collaborator's (Jack's) CSI fingerprints are used as the condition to generate Alice's real-time CSI fingerprints, and Eve transmits pilot signals to Bob to impersonate Alice via spoofing attacks.}
    \label{Authenication}
\end{figure}

As illustrated in Figure \ref{Authenication}, we consider a dynamic scenario, which is composed of the nodes below.

\begin{itemize}
    \item \textit{Alice:}\hspace{0.3em}Normal devices that need to be correctly identified.
    \item \textit{Jack:}\hspace{0.3em}Collaborative devices which have been correctly identified and can be used to assist the identification of Alice nearby.
    \item \textit{Bob:}\hspace{0.3em}Nodes used to identify IoT devices, such as base stations (BSs).
    \item \textit{Eve:}\hspace{0.3em}In dynamic scenarios, an eavesdropper can impersonate Alice.
\end{itemize}
%原来的如下，是移动场景，修改为动态场景了。
% In the mobile scenario, we assume that Alice and Jack move in the same direction at the same velocity, meaning they remain stationary relative to each other. The distance between Alice and Jack is denoted as $d$. Jack is the collaborative device that has already been correctly identified. Alice is the unknown device that needs to be identified. Since Alice and Jack remain at a fixed and close distance, a stable relationship between their CSI fingerprints can be ensured through channel extrapolation in the user domain \cite{r34}. Bob is the receiver that covers the communication range of both Alice and Jack. The distance between Bob and Alice is denoted as $D$, where $d \ll D$. Eve is a spoofing attacker within the coverage area of Bob, capable of impersonating Alice to attack Bob, thus requiring identity authentication to mitigate this threat.

\begin{table*}[htbp]
\caption{Summary of Key Notations and Parameters}
\label{tab:notations}
\centering
% 设置颜色环境
% \color{blue}           % 将表格内的文字设为蓝色
% \arrayrulecolor{blue}  % 将表格的横线和竖线设为蓝色

% \textwidth 让表格撑满双栏的总宽度
% @{\extracolsep{\fill}} 自动计算列间距以填满宽度
\setlength{\tabcolsep}{18pt} % 设置表格列之间的水平距离为10pt
\begin{tabular}{|l|l|l|l|}
% \begin{tabular*}{\textwidth}{|l|l|l|l|}
\hline
\textbf{Notation} & \textbf{Definition} & \textbf{Notation} & \textbf{Definition} \\ \hline
$d$ & Fixed distance between Alice and Jack & $M$ & Number of antennas (Alice/Jack) \\ \hline
$f$ & Carrier frequency & $L$ & Number of multipath paths \\ \hline
$K$ & Total number of OFDM subcarriers & $y_A, y_J$ & Signal received by Bob from Alice and Jack\\ \hline
$n(t)$ & Channel noise vector & $h_{A}, h_{J}$ & True CSI of Alice and Jack \\ \hline
$\hat{h}_{A}, \hat{h}_{J}$ & Estimated CSI fingerprints & $\tilde{h}_A$ & Generated CSI of Alice \\ \hline
$\mathcal{D}(\cdot)$ & Distance metric for authentication & $k$ & Attack frequency ratio of Eve \\ \hline
$\nu_{Alice}, \nu_{Eve}$ & Alice's emission and Eve's attack frequency & $T$ & Total number of diffusion steps \\ \hline
$m$ & Binary mask for diffusion & $\alpha_t, \beta_t$ & Noise schedule hyperparameters \\ \hline
$\tilde{h}_t$ & Noisy concatenated CSI at step $t$ & $h_t$ & Noisy CSI of Alice at step $t$ \\ \hline
% $\{Q, K, V\}$ & Cross-Attention matrices & $\nu_{Alice}$ & Alice's signal emission frequency \\ \hline
% \end{tabular*}
\end{tabular}
\arrayrulecolor{black} % 这是一个好习惯：表格结束后将线条颜色恢复为黑色，以免影响后续表格
\end{table*}

In the dynamic scenario, we assume that both Alice and Jack are stationary while the surrounding environment varies rapidly. The distance between Alice and Jack is denoted as $d$, and remains fixed during the process. Jack is the collaborative device that has already been correctly identified, while Alice is the unknown device requiring identification. This fixed and close distance allows a stable relationship between their CSI fingerprints to be established through channel extrapolation in the user domain \cite{r34}. Bob is the receiver covering the communication range of both Alice and Jack. The distance between Bob and Alice is denoted as $D$, where $d \ll D$. Eve is a spoofing attacker within Bob’s coverage area who may impersonate Alice. For clarity, the main symbols and parameters used in this paper are summarized in Table \ref{tab:notations}.

% $f_{Alice}$ and $f_{Eve}$ represent Alice's signal emission frequency and Eve's attack frequency, respectively.

\subsection{Channel Model}
We consider an uplink authentication process, where transmitters are equipped with \textit{M} antennas in the form of uniform linear array (ULA), and Bob is equipped with \textit{N} antennas in the same form. Let ${h}_{m,n}$ denote the channel from the transmitter's ${m}^{th}$ antenna to Bob's ${{n}^{th}}$ antenna at the frequency \textit{f}. Then, the multipath channel model is expressed as
\begin{equation}
\label{h_{m,n}}
    h_{m,n}(f,\tau) = \sum\limits_{l=1}^L a_le^{-j2\pi f\tau_l + j\varphi_l}\delta(\tau-\tau_l),
\end{equation}

\noindent where \textit{L} presents the number of different transmission paths. $a_l$ is the path gain, which is based on the frequency \textit{f}, the distance from Alice to Bob $d$ and the antenna gain. The phase $\varphi_l$ also depends on the material of the scatterer and the wave incident angle at the scatterer. $\tau_l$ is a frequency-independent time delay of the path $l$. In the wide-band channel, the coefficient of the $n^{th}$ subcarrier in an orthogonal frequency division multiplexing (OFDM) system is expressed as
\begin{equation}
\label{H_{m,n}}
    H_{m,n}(f, k) = \sum\limits_{l=1}^L a_le^{-j2\pi f\tau_l + j\varphi_l}e^{-j2\pi \tau_l n/K},
\end{equation}

\noindent where $f_k$ is the frequency of the $n^{th}$ subcarrier relative to the center frequency $f$. $K$ represents the total number of the subcarriers. We define the $M\times K$ channel vector $H(f) = [H_{1,n}(f),\ldots,H_{M,n}(f)]$, where $H_{m,n}(f) = [H_{m,n}(f,0),\ldots,H_{m,n}(f,K-1)]$. Furthermore, the signal passed from Alice to Bob at time slot $t$ is expressed as
\begin{equation}
\label{s_{t}}
    y_A(t) = h_A(f,t)\cdot x(t)+ n_A(t),
\end{equation}

\noindent where $y_A\left(t\right)$ is the signal that Bob received, and $x(t)$ represents the guide signal transmitted from the transmitter to Bob at time slot $t$. $n_A(t)\sim \mathcal{C}\mathcal{N}(0, \sigma_A^2\textbf{\textit{I}})$ is the channel noise, assumed to be a complex Gaussian random variable with zero mean and variance $\sigma_A^2$. We employ the Least Squares (LS)-based channel estimation, and the channel $h$ at time slot $t$ is expressed as

\begin{equation}
\label{{h}_A(f,t)}
    \hat{h}_A(f,t) = \left( x(t)^H h(t) \right)^{-1} x(t)^H y_A(t),
\end{equation}

\noindent where $ x(t)^H$ represents the Hermitian transpose of $x\left(t\right)$. $\hat{h}_A(f,t)$ denotes Alice's CSI fingerprints.
The signal Bob received from Jack is denoted as
\begin{equation}
\label{s_y{t}}
    y_J(t) = h_J(f,t)\cdot x(t)+ n_J(t),
\end{equation}

\noindent where $x(t)$ represents the guide signal transmitted from Jack to Bob at time slot $t$. $n_J(t)\sim \mathcal{C}\mathcal{N}(0, \sigma_J^2\textbf{\textit{I}})$ is the channel noise, assumed to be a complex Gaussian random variable with zero mean and variance $\sigma_j^2$. Correspondingly, the channel estimation result $\hat{h}_J(f,t)$ is expressed as
\begin{equation}
\label{{h}_J(f,t)}
    \hat{h}_J(f,t) = \left( x(t)^H x(t) \right)^{-1} x(t)^H y_J(t).
\end{equation}

According to channel extrapolation in user domain \cite{r27,r28}, we know that neighboring users share similar transmission paths and surrounding environments. Thus, Alice experiences the same channel noise as that of the neighboring device Jack during time slot $t$, that is, $n_A(t)= n_J(t)$ and $\sigma_A^2=\sigma_J^2$. The relationship between the channel of Alice $h_A\left(t\right)$ and that of Jack $h_J\left(t\right)$ can be expressed as
 \begin{equation}
\label{{f(x,y)}}
    h_A(t) = \varphi(h_J(t)), 
\end{equation}

\noindent where $\varphi(\cdot)$ is the mapping relationship from $h_A\left(t\right)$ to $h_J\left(t\right)$.

\begin{algorithm}[t]
\caption{The Steps of the Proposed APEG Scheme}
\label{algorithm1}
\begin{algorithmic}[1]
\State \textbf{Training Phase:}
\State Collect CSI fingerprint samples of Alice $\hat{h}_{A,t}$ and those of Jack $\hat{h}_{J,t}$ at different time slots $t$.
\State Convert the complex CSI fingerprint matrix $\hat{h}_{A,t}$ and $\hat{h }_{J,t}$ into 2-channel images.
\State Use \textbf{Algorithm \ref{algorithm2}} or \textbf{Algorithm \ref{algorithm3}} to train the GAI model, which can generate Alice's CSI fingerprint $\tilde{h}_{A,t}$ using $\hat{h}_{J,t}$.
% \State Sample Eves' CSI fingerprints $\hat{h}_t^E$ around Alice in the validation set according to the attack frequency $K$.
% \For{each sampled Alice $m=1,2,...,M$}
%     \State Generate Alice's CSI fingerprint $\tilde{h}_{A,m}$ at Bob using $\hat{h}_{J,m}$.
%     \For{each sampled Eve $k=1,2,...,K$}
%     \If{$\mathcal{D}(\tilde{h}_{A,m}, \hat{h}_{A,m})>\mathcal{D}(\tilde{h}_{A,m}, \hat{h}_{E,m,k})$}
%         \State $\mathbb{I}_{\text{auth}}(m)=1$.
%     \Else{}
%         \State $\mathbb{I}_{\text{auth}}(m)=0, \gamma_{m,k} = \mathcal{D}(\tilde{h}_{A,m}, \hat{h}_{E,m,k}) $
%     \EndIf
%     \EndFor
% \EndFor
% \If{$\Gamma_{M,K}$ is a non-zero set}
%     \State Select the maximum $\gamma = \gamma_{m,k}$ from $\Gamma_{M,K}$ as Authentication threshold.
% \Else{}
%     \State Select the minimum $\gamma = \mathcal{D}(\tilde{h}_{A,m}, \hat{h}_{A,m})$ as Authentication threshold.
% \EndIf
\State \textbf{Authentication Phase:}
\For{each time slot $t=1,2,\ldots, T$}
    \State Alice and Eves send signals to Bob for authentication.
    \State Bob receives $s$ CSI fingerprint samples from Alice and Eves.
    \State Bob obtains Jack's CSI fingerprint $\hat{h}_{J,t}$.
    \State By using (\ref{{f(x,y)}}) with the trained GAI model, Bob obtains the predicted Alice's CSI fingerprint $\tilde{h}_{A,t}$. 
    \State Bob obtains authentication results based on (\ref{pla_dis}).
\EndFor   
% \State Bob receives the pilot signal from Jack and performs channel estimation using~\eqref{{h}_J(f,t)} to obtain $\hat{h}_{J,t}$ at time slot $t$.
% \State Bob uses the trained GAI model to generate Alice's CSI fingerprint $\tilde{h}_{A,t}$ at time slot $t$ with the input $\hat{h}_{J,t}$.
% \State Bob receives the pilot signal from Unverified device and obtains the CSI fingerprint $h_{A\lor E,t}$.
% \State Sample Eves' CSI fingerprints $\hat{h}_t^E$ around Alice in the validation set according to the attack frequency $K$.
% \For{each sampled Alice $m=1,2,...,M$}
%     \State Generate Alice's CSI fingerprint $\tilde{h}_{A,m}$ at Bob using $\hat{h}_{J,m}$.
%     \For{each sampled Eve $k=1,2,...,K$}
%     \If{$\mathcal{D}(\tilde{h}_{A,m}, h_{A\lor E,m,k})<\mathcal{D}(\tilde{h}_{A,m}, h_{A\lor E,m,k+1}$)}
%         \State $\mathbb{I}_{\text{auth}}(m)=1$.
%         \State $\hat h_{A,m} = h_{A\lor E,m,k}$
%         \State $h_{A\lor E,m,k+1}=h_{A\lor E,m,k}$
%     \Else{}
%         \State $\mathbb{I}_{\text{auth}}(m)=0 $
%     \EndIf
%     \EndFor
%     \State Output $\hat h_{A,m}$ and corresponding Alice device $m$.
% \EndFor
% 修改2
% \If{$\Gamma_{M,K}$ is a non-zero set}
%     \State Select the maximum $\gamma = \gamma_{m,k}$ from $\Gamma_{M,K}$ as Authentication threshold.
% \Else{}
%     \State Select the minimum $\gamma = \mathcal{D}(\tilde{h}_{A,m}, \hat{h}_{A,m})$ as Authentication threshold.
% \EndIf
% \If{$\mathcal{D}(\tilde{h}_{A,t}, \hat{h}_{A\lor E,t)})>\gamma$}
%     \State $\mathbb{I}_{\text{auth}}(t)=1$.
% \Else{}
%         \State $\mathbb{I}_{\text{auth}}(t)=0$
% \EndIf
\end{algorithmic}
\end{algorithm}

\subsection{Proposed Adaptive PLA Scheme With Channel Extrapolation and GAI (APEG)}

The complete procedure of the proposed APEG scheme is detailed in Algorithm \ref{algorithm1}, which includes training and authentication phases as follows.
\subsubsection{Training Phase}
Bob learns the features of the dynamic Alice's estimated CSI fingerprints $\hat h_A$ and the known estimated CSI fingerprints of its neighboring Jack $\hat h_J$. According to the principle of channel extrapolation in the user domain, there exists spatial correlation between $\hat h_A$ and $\hat h_J$. We train a GAI model to capture the joint CSI fingerprint representation between $\hat h_A$ and $\hat h_J$. The training dataset can be constructed by collecting CSI fingerprints of Alice and Jack at different locations via Bob. 
% \item In the authentication phase, we define $\mathcal{D}(X, Y)$ to represent the distance between $X$ and $Y$ across multiple evaluation metrics and $\mathbb{{I}}_{\text{auth}}(\cdot)$ to represent if the authentication is correct. Moreover, we assume that Eve attacks with a frequency $K$ and is randomly distributed within a radius $r$ around Alice. Bob estimates the CSI fingerprints of Jack, Alice, and Eve separately. Since Jack is a pre-authenticated collaborative device, the features derived from $\hat h_J$ are used to discriminate between the channel estimates of Alice and Eve through the distance function $\mathcal{D}(\tilde{h}_{A,t}, \hat{h}_{A\lor E,t})$. Finally, Alice is authenticated by selecting the minimum distance from the compared distance values.
\subsubsection{Authentication Phase}
We predict Alice's CSI fingerprint $\tilde h_{A,t}$ by using the trained GAI model with $\hat h_{J,t}$. Then, we define $\mathcal{D}(\cdot)$ to represent the distance between $\tilde h_{A,t}$ and $\hat h_{t}$ across multiple evaluation metrics, such as Structural Similarity Index (SSIM), Peak Signal-to-Noise Ratio (PSNR), Cosine similarity, and Normalized Mean Squared Error (NMSE).
We further have

\begin{equation}
\label{pla_dis}
  \mathcal{D}(A)<\mathcal{D}(E),
\end{equation}

\noindent where $\mathcal{D}(A)$ represents the distance between and $\tilde h_{A,t}$ and $\hat h_{A,t}$, and $\mathcal{D}(E)$ represents the distance between and $\tilde h_{A,t}$ and $\hat h_{E,t}$.
In the authentication phase, there are $s$ CSI fingerprint samples, including Alice's CSI fingerprints and Eve's cheating fingerprints. Sorting them according to a numerical value, we can obtain

\begin{equation}
\label{pla_order}
  \mathcal{D}(\pi(1)) < \mathcal{D}(\pi(2)) < \cdots < \mathcal{D}(\pi(s)),
\end{equation}
where \(\pi(1), \pi(2), \ldots, \pi(s)\) represent the indices sorted in ascending order based on \(\mathcal{D}(t)\). Then, based on (\ref{pla_dis}), the set \(\{\pi(1), \pi(2), \ldots, \pi(ks)\}\) corresponds to the $ks$ top Alice devices with the smallest distance values, while the remaining set \(\{\pi(ks+1), \pi(ks+2), \ldots, \pi(s)\}\) corresponds to the Eves. The ratio between Alice's emission frequency and Eve's attack frequency $k$ ($k\in[0,1]$) is expressed as

\begin{equation}
\label{pla_k}
  k = \frac{\nu_{Alice}}{\nu_{Alice}+\nu_{Eve}},
\end{equation}
where $\nu_{Alice}$ and $\nu_{Eve}$ represent Alice's signal emission frequency and Eve's attack frequency, respectively.

To operationalize the proposed APEG scheme in dynamic environments, such as Industrial IoT (IIoT) scenarios, Algorithm \ref{algorithm1} is mapped to a continuous workflow. Initially, the offline phase (Steps 2--4) establishes the baseline spatial correlation, utilizing the collaborator (Jack) as an ``environmental anchor'' to learn the mapping from historical data. Subsequently, the online authentication phase (Steps 6--11) executes periodically: Bob receives uplink pilots and acquires real-time CSI. Using Jack's CSI as the anchor, Bob extrapolates Alice's theoretical fingerprint via the trained model and performs the final decision based on the similarity ranking logic as step 11. Finally, to counteract environmental drift, high-confidence samples from Step 11 are selectively utilized for online fine-tuning (via Step 4), ensuring the reliability of continuous authentication.
% After training, the GAI model uses Jack's CSI fingerprint $\hat h_J$ to generate Alice's fingerprint $\tilde h_A$. We define $\mathcal{D}(X, Y)$ to represent the distance between $X$ and $Y$ across multiple evaluation metrics, and use this distance to determine the authentication threshold $\gamma_{m,k}$ to distinguish Alice and Eve. Moreover, we define $\mathbb{{I}}_{\text{auth}}(\cdot)$ to represent if the authentication is correct. 

\begin{figure}[t]
    \centering
    \includegraphics[width=0.4\textwidth,trim=0 0 0 0,clip]{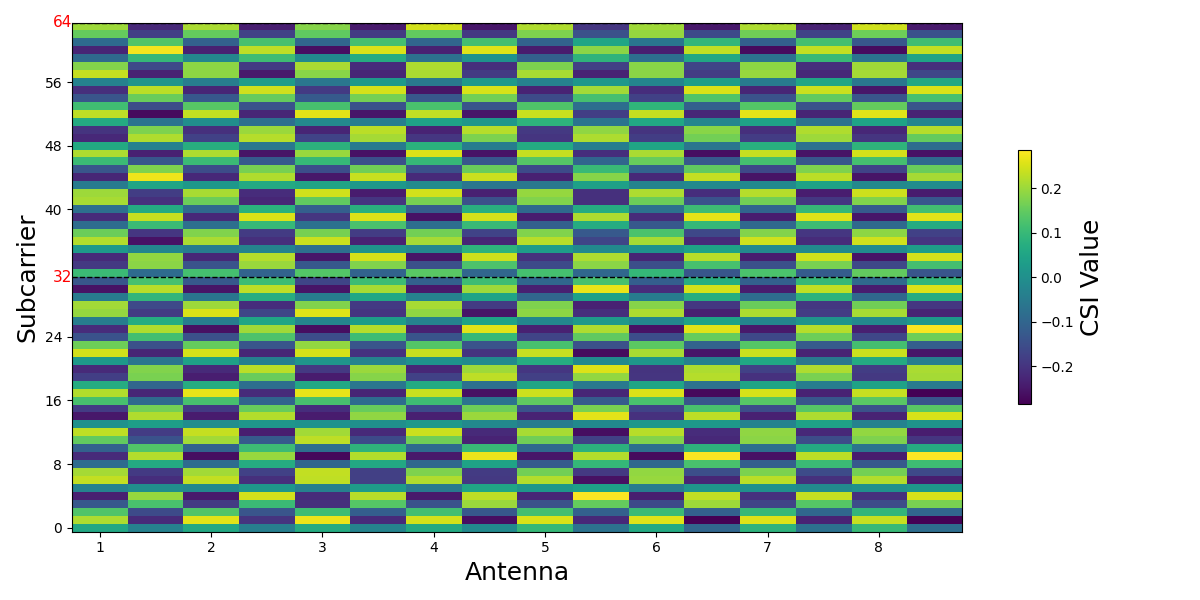}
    \caption{Visualization of the real parts of Alice’s and Jack’s CSI fingerprints generated from the DeepMIMO 'O1' dataset \cite{Alkhateeb2019}, vertically concatenated and displayed as a heatmap. The lower half shows Alice’s fingerprints and the upper half shows Jack’s fingerprints.}
    \label{1893_2074_real_image}
\end{figure}
% The lower half shows Alice’s fingerprint and the upper half shows Jack’s fingerprint.

\section{Proposed DM-based CSI Fingerprint Generation Schemes}
% diffusion
\subsection{CSI Fingerprints as Images}
We construct two images of identical dimensions from Alice’s CSI fingerprint and the corresponding Jack's CSI fingerprint at the same time. Each image has a size of $M\times N$, where $N$ is the number of subcarriers and $M$ represents the number of antennas. To bridge the gap between the complex-valued channel $\mathbf{H} \in \mathbb{C}^{M \times N}$ and standard real-valued deep learning architectures, we designed two channels in each image: the first stores the real part, and the second stores the imaginary part. This mapping transforms the data into a real-valued tensor $\mathcal{X} \in \mathbb{R}^{2 \times M \times N}$. While similar image-based processing has been successfully applied in the angle-delay domain \cite{wen2018deep}, we extend this paradigm to the frequency domain. This allows us to leverage 2D convolutions to capture structural correlations across antennas and subcarriers while preserving the full-resolution phase information inherent in the raw channel. For example, the image of the first channel is illustrated in Figure \ref{1893_2074_real_image}.

\begin{rem}
\upshape
\textbf{Justification for Domain Selection:} It is worth noting that while the Angle-Delay Domain (ADD) offers sparse representations that physically align with multipath propagation, this paper primarily targets the frequency-domain CSI for the two reasons. First, the frequency domain represents the raw channel response, preserving the full-resolution amplitude and phase information essential for distinguishing subtle identity fingerprints. Second, transforming to the ADD typically involves Discrete Fourier Transforms (DFT) or compressive sensing, which introduce spectral leakage and quantization errors due to off-grid multipath components. These artifacts can degrade the fine-grained features required for high-precision authentication. Thus, we adopt the frequency domain to maximize fingerprint fidelity. 
% leaving the exploration of ADD-based structural robustness for future extensions.
\end{rem}

\subsection{Proposed CCMDM-based Fingerprint Generation Scheme}
As a class of GAI techniques, DM aims at constructing a forward diffusion process that gradually adds noise to the original data, transforming it into an approximately isotropic Gaussian distribution \cite{cao2024survey}. Thus, we employ CCMDM to iteratively recover the original data distribution from the noise, thereby enabling the modeling of complex data generation.

We assume that Alice's CSI fingerprints $h_0$ follows a prior distribution ${h_0} \sim q\left( {{h_0}} \right)$, where $h_0=\hat{h}_A$, and a reverse diffusion process ${p_\theta }\left( {{h_{0:T}}} \right)$ is defined conditioned on Alice's CSI fingerprints. Since CCMDM is employed, we concatenate Alice's and Jack's CSI fingerprints along the last dimension to form the input, denoted as ${\tilde h_0} = \left[ {{h_0}\left\| {{\hat{h}_J}} \right.} \right]$. Gaussian noises are added only to the region corresponding to Alice, while the region corresponding to Jack remains clean. Accordingly, the joint probability distribution of the reverse generation process can be expressed as ${p_\theta }\left( {{{\tilde h}_{0:T}}} \right)$. Then, based on \cite{ho2020denoising}, the corresponding latent variable model is formulated as

\begin{equation}
\label{p1}
   {p_\theta }\left( {{{\tilde h}_0}} \right) = \int {{p_\theta }\left( {{{\tilde h}_0},{h_{1:T}}} \right)d} {h_{1:T}},
\end{equation}
where ${h_1},\ldots,{h_T}$ are latent variables with the same dimensionality as $h_0$, representing Alice's CSI fingerprints corrupted by noises at time step $t$. Furthermore, ${p_\theta }\left( {{h_{0:T}}} \right)$ is defined as a Markov chain starting from the initial state $p\left( h_T \right) = \mathcal{N}\left( h_T; 0, \textbf{\textit{I}} \right)$, where each transition is modeled by a learned Gaussian conditional distribution. The reverse process in DMs can thus be further formulated as

\begin{equation}
\label{p2}
    {p_\theta }\left( {{h_{0:T}}} \right) = p\left( {{h_T}} \right)\prod\limits_{t = 1}^T {{p_\theta }\left( {{h_{t - 1}}\left| {{h_t}} \right.} \right)},
\end{equation}

\noindent where ${p_\theta }\left( {{h_{t - 1}}\left| {{h_t}} \right.} \right)$ is the conditional probability distribution of generating the sample $h_{t-1}$ at time step $t-1$, given Alice's noisy CSI fingerprints $h_{t}$ at time step $t$. It is defined as

\begin{equation}
\label{p3}
    {p_\theta }\left( {{h_{t - 1}}\left| {{h_t}} \right.} \right) = \mathcal{N}\left( {{h_{t - 1}};{\mu _\theta }\left( {{h_t},t} \right),\sum\nolimits_\theta  {\left( {{h_t},t} \right)} } \right).
\end{equation}

The reverse process follows a Markov chain that progressively denoises the Gaussian state to recover the probability distribution of Alice's original CSI fingerprints $h_0$. However, the parameters of the conditional Gaussian noise used in this process must be learned progressively during the forward process. Therefore, the forward process is also modeled as a Markov chain as

\begin{equation}
\label{p4}
    q\left( {{h_{1:T}}\left| {{h_0}} \right.} \right) = \prod\limits_{t = 1}^T {q\left( {{h_t}\left| {{h_{t - 1}}} \right.} \right)} ,
\end{equation}

\begin{figure*}[!b]
    \hrulefill
    \normalsize
    \setcounter{equation}{14}
    \begin{equation}
    \label{p5}
        q\left( {{{\tilde h}_t}\left| {{{\tilde h}_{t - 1}}} \right.} \right) = \mathcal{N}\left( {{{\tilde h}_t};\left( {1 - m} \right) \cdot {{\tilde h}_t} + m \cdot \sqrt {1 - {\beta _t}} {{\tilde h}_{t - 1}},m \cdot {\beta _t}\textbf{\textit{I}}} \right)
    \end{equation}
    \setcounter{equation}{15}
    \begin{equation}
    \label{p6}
        q\left( {{{\tilde h}_t}\left| {{{\tilde h}_0}} \right.} \right) = \mathcal{N}\left( {{{\tilde h}_t};\left( {1 - m} \right) \cdot {{\tilde h}_0} + m \cdot \sqrt {{{\bar \alpha }_t}} {{\tilde h}_0},m \cdot \left( {1 - {{\bar \alpha }_t}} \right)\textbf{\textit{I}}} \right)
    \end{equation}
\end{figure*}

\noindent where $q\left( {{h_{1:T}}\left| {{h_0}} \right.} \right)$ represents the approximate posterior of the complete diffusion process. Since noise is added only to the region corresponding to Alice at each diffusion step, the Markov forward process is constructed as~\eqref{p5},
% \begin{equation}
% \label{p5}
%     q\left( {{{\tilde h}_t}\left| {{{\tilde h}_{t - 1}}} \right.} \right) = N\left( {{{\tilde h}_t};\left( {1 - m} \right) \cdot {{\tilde h}_t} + m \cdot \sqrt {1 - {\beta _t}} {{\tilde h}_{t - 1}},m \cdot {\beta _t}I} \right)
% \end{equation}
where $\beta_t$ is the noise added at time step $t$, mask $m = \left[ {{1_A},{0_J}} \right]$, ${h_t} = m \cdot {\tilde h_t}$, and ${\hat{h}_J} = \left( {1 - m} \right) \cdot {\tilde h_t}$. Since the forward process allows sampling $h_t$ at any time step $t$ in a closed-form expression, we introduced simplified notation by defining ${\alpha _t} = 1 - {\beta _t}$ and ${\bar \alpha _t} = \prod\nolimits_{i = 1}^t {{\alpha _i}}$. The closed-form expression of the forward process at arbitrary time step $t$ is given in~\eqref{p6}.
% \begin{equation}
% \label{p6}
%     q\left( {{{\tilde h}_t}\left| {{{\tilde h}_0}} \right.} \right) = N\left( {{{\tilde h}_t};\left( {1 - m} \right) \cdot {{\tilde h}_0} + m \cdot \sqrt {{{\bar \alpha }_t}} {{\tilde h}_0},m \cdot \left( {1 - {{\bar \alpha }_t}} \right)I} \right)
% \end{equation}
The posterior mean of the forward process is parameterized as 

\begin{equation}
\label{p7}
    {\tilde h_t}\left( {{h_0},\varsigma } \right) = \left( {1 - \left( {1 - \sqrt {{{\bar \alpha }_t}} } \right)m} \right) \cdot {\tilde h_0} + m \cdot \sqrt {1 - {{\bar \alpha }_t}} \varsigma ,
\end{equation}

% \begin{figure}[t]
%     \centering
%     \includegraphics[width=0.5\textwidth,trim=0 0 0 0,clip]{Network_system.png}
%     \caption{Authentication performance of different schemes based on SSIM}
%     \label{fig2:"Authentication performance of different schemes based on SSIM"}
% \end{figure}

\noindent where $\varsigma  \sim \mathcal{N}\left( {0,\textbf{\textit{I}}} \right)$. With the forward and reverse processes formally established in~(\ref{p2})--(\ref{p6}), the objective of the model is to learn the reverse denoising process to reconstruct Alice's fingerprint $\hat h_0$. Therefore, we aim to maximize the log-likelihood of the original data $\mathop {\max }\limits_\theta  \log {p_\theta }\left( {{h_0}} \right)$, which leads to the construction of a variational lower bound (ELBO) as

\begin{equation}
\label{p8}
    \mathcal{L} = {\mathbb{E}_q}\left[ { - \log p\left( {{h_T}} \right) - \sum\limits_{t \ge 1} {\log \frac{{{p_\theta }\left( {{h_{t - 1}}\left| {{h_t}} \right.} \right)}}{{q\left( {{h_{1:T}}\left| {{h_0}} \right.} \right)}}} } \right] .
\end{equation}

By decomposing (\ref{p8}), we observe that the ultimate objective is to minimize the distance between $q\left( {{h_{t - 1}}\left| {{h_t},{h_0}} \right.} \right)$ and ${p_\theta }\left( {{h_{t - 1}}\left| {{h_t}} \right.} \right)$, which encourages the learned reverse process to approximate the true reverse trajectory as closely as possible. This objective is represented using the Kullback–Leibler (KL) divergence, denoted as $KL\left( {q\left( {{h_{t - 1}}\left| {{h_t},{h_0}} \right.} \right)\left\| {{p_\theta }\left( {{h_{t - 1}}\left| {{h_t}} \right.} \right)} \right.} \right)$. Moreover, since the sequence ${h_0} \to {h_1} \to \ldots \to {h_{t - 1}} \to {h_t}$ forms a linear Gaussian chain, the closed-form solution can be directly derived according to

\begin{equation}
\label{p9}
    q\left( {{h_{t - 1}}\left| {{h_t},{h_0}} \right.} \right) = \mathcal{N}\left( {{h_{t - 1}};{{\tilde \mu }_t}\left( {{h_t},{h_0}} \right),{{\tilde \beta }_t}\textbf{\textit{I}}} \right),
\end{equation}
where
\begin{equation}
\label{p10}
    {\tilde \mu _t}\left( {{h_t},{h_0}} \right) = \frac{{\sqrt {{{\bar \alpha }_{t - 1}}} {\beta _t}}}{{1 - {{\bar \alpha }_t}}}{h_0} + \frac{{\sqrt {{\alpha _t}} \left( {1 - {{\bar \alpha }_{t - 1}}} \right)}}{{1 - {{\bar \alpha }_t}}}{h_t}
\end{equation}
and
\begin{equation}
\label{p11}
    {\tilde \beta _t} = \frac{{1 - {{\bar \alpha }_{t - 1}}}}{{1 - {{\bar \alpha }_t}}}{\beta _t} .
\end{equation}

Then, to minimize $KL\left( {q\left( {{h_{t - 1}}\left| {{h_t},{h_0}} \right.} \right)\left\| {{p_\theta }\left( {{h_{t - 1}}\left| {{h_t}} \right.} \right)} \right.} \right)$, we reparameter the loss function by ${p_\theta }\left( {{h_{t - 1}}\left| {{h_t}} \right.} \right) = \mathcal{N}\left( {{h_{t - 1}};{\mu _\theta }\left( {{h_t},t} \right),\sum\nolimits_\theta  {\left( {{h_t},t} \right)} } \right)$ as

\begin{equation}
\label{p12}
    \mathcal{L} = {\mathbb{E}_q}\left[ {\frac{1}{{2{\sigma ^2}}}{{\left\| {{{\tilde \mu }_t}\left( {{h_t},{h_0}} \right) - {\mu _\theta }\left( {{h_t},t} \right)} \right\|}^2}} \right] + C ,
\end{equation}

% \begin{figure*}[b]
%     \hrulefill
%     \normalsize
%     \setcounter{equation}{19}
%     \begin{equation}
%     \label{p16}
%         {\tilde h_{t - 1}} = \left( {1 - m} \right) \cdot {\tilde h_t} + m \cdot \left[ {\frac{1}{{\sqrt {{\alpha _t}} }}\left( {{{\tilde h}_t} - \frac{{{\beta _t}}}{{\sqrt {1 - {{\bar \alpha }_t}} }}{\varsigma _\theta }\left( {m \cdot {{\tilde h}_t},t} \right)} \right) + {{\tilde \beta }_t}{z_t}} \right]
%     \end{equation}
%     \setcounter{equation}{20}
%     \begin{equation}
%     \label{p13}
%         L = {E_q}\left[ {\frac{1}{{2{\sigma ^2}}}{{\left\| {{{\tilde \mu }_t}\left( {{h_t}\left( {{h_0},\varsigma } \right),\frac{{\rm{1}}}{{\sqrt {{{\bar \alpha }_t}} }}\left( {{h_t}\left( {{h_0},\varsigma } \right) - \sqrt {1 - {{\bar \alpha }_t}} \varsigma } \right)} \right) - {\mu _\theta }\left( {{h_t},t} \right)} \right\|}^2}} \right] + C
%     \end{equation}
%     \setcounter{equation}{23}
%     \begin{equation}
%     \label{p15}
%         {L_{DM}}\left( \theta  \right) = {{\rm E}_{{h_0},t,{\varsigma _t}}}\left[ {\left\| {m \cdot \left( {{\varsigma _t} - {\varsigma _\theta }\left( {\left( {1 - \left( {1 - \sqrt {{{\bar \alpha }_t}} } \right)m} \right) \cdot {{\tilde h}_0} + m \cdot \sqrt {1 - {{\bar \alpha }_t}} {\varsigma _t},t} \right)} \right)} \right\|_2^2} \right]
%     \end{equation}
% \end{figure*}

\begin{algorithm}[t]
\caption{The Steps of the Proposed CCMDM-based Fingerprint Generation Scheme}
\label{algorithm2}
\begin{algorithmic}[1]
\State \textbf{Training Phase}
\State Take Alice's fingerprint $h_0$ and Jack's fingerprint $\hat{h}_J$ as a 2-channel image $\tilde h_0 =  \left[ {{h_0}\left\| {{\hat{h}_J}} \right.} \right]$
\Repeat{}
\State Sample Alice’s CSI fingerprint: ${h_0} \sim q(h_0)$
\State Sample diffusion step: $t \sim \text{Uniform}(1,\ldots,T)$
\State Sample noise: $\varsigma \sim \mathcal{N}(0,\textbf{\textit{I}})$
\State Perform gradient descent computation at each step $\nabla_\theta \left\| \varsigma -  {\varsigma _\theta }\left( {\left( {1 - \left( {1 - \sqrt {{{\bar \alpha }_t}} } \right)m} \right) \cdot {{\tilde h}_0} + m \cdot \sqrt {1 - {{\bar \alpha }_t}} {\varsigma},t} \right)\right \|^2$
\Until{\eqref{p15} converges}
\State \textbf{Fingerprint Generating Phase}
\State Input: $\tilde h_T =  \left[ {{h_T}\left\| {{\hat{h}_J}} \right.} \right]$, where $h_T \sim \mathcal{N}(0,\textbf{\textit{I}})$
\For{$t=T,\ldots,1$}
    \State Sample perturbation noise $z_t \sim \mathcal{N}(0,\textbf{\textit{I}})$ (only if $t > 1$)
    \State Then, perform a step-by-step denoising process on $\tilde h_T$ depending on the noise predicted by~\eqref{p16}
\EndFor
\State return $\left[ {{\tilde h_A}\left\| {{\hat{h}_J}} \right.} \right]$, where $\tilde{h}_A$ is the generated channel of Alice. \end{algorithmic}
\end{algorithm}

\noindent where $\sum\nolimits_\theta  {\left( {{h_t},t} \right)}  = {\sigma ^2}I$ and $C$ is a constant value. By substituting~\eqref{p7}, $L$ can be further expressed as~\eqref{p13}.
To ensure the loss converges downward, we represent ${\mu _\theta }\left( {{h_t},t} \right)$ as 

\begin{figure*}[!b]
    \hrulefill
    \normalsize
    \setcounter{equation}{22}
    \begin{equation}
    \label{p13}
        \mathcal{L} = {\mathbb{E}_q}\left[ {\frac{1}{{2{\sigma ^2}}}{{\left\| {{{\tilde \mu }_t}\left( {{h_t}\left( {{h_0},\varsigma } \right),\frac{{\rm{1}}}{{\sqrt {{{\bar \alpha }_t}} }}\left( {{h_t}\left( {{h_0},\varsigma } \right) - \sqrt {1 - {{\bar \alpha }_t}} \varsigma } \right)} \right) - {\mu _\theta }\left( {{h_t},t} \right)} \right\|}^2}} \right] + C
    \end{equation}
    \setcounter{equation}{25}
    \begin{equation}
    \label{p15}
        {\mathcal{L}_{DM}}\left( \theta  \right) = {{\rm \mathbb{E}}_{{h_0},t,{\varsigma _t}}}\left[ {\left\| {m \cdot \left( {{\varsigma _t} - {\varsigma _\theta }\left( {\left( {1 - \left( {1 - \sqrt {{{\bar \alpha }_t}} } \right)m} \right) \cdot {{\tilde h}_0} + m \cdot \sqrt {1 - {{\bar \alpha }_t}} {\varsigma _t},t} \right)} \right)} \right\|_2^2} \right]
    \end{equation}
    \begin{equation}
    \label{p16}
        {\tilde h_{t - 1}} = \left( {1 - m} \right) \cdot {\tilde h_t} + m \cdot \left[ {\frac{1}{{\sqrt {{\alpha _t}} }}\left( {{{\tilde h}_t} - \frac{{{\beta _t}}}{{\sqrt {1 - {{\bar \alpha }_t}} }}{\varsigma _\theta }\left( {m \cdot {{\tilde h}_t},t} \right)} \right) + {{\tilde \beta }_t}{z_t}} \right]
    \end{equation}
    \setcounter{equation}{27}
\end{figure*}

% \begin{equation}
% \label{p13}
%     L = {E_q}\left[ {\frac{1}{{2{\sigma ^2}}}{{\left\| {{{\tilde \mu }_t}\left( {{h_t}\left( {{h_0},\varsigma } \right),\frac{{\rm{1}}}{{\sqrt {{{\bar \alpha }_t}} }}\left( {{h_t}\left( {{h_0},\varsigma } \right) - \sqrt {1 - {{\bar \alpha }_t}} \varsigma } \right)} \right) - {\mu _\theta }\left( {{h_t},t} \right)} \right\|}^2}} \right] + C
% \end{equation}

\setcounter{equation}{23}
\begin{equation}
\label{p13}
    {\mu _\theta }\left( {{h_t},t} \right) = \frac{1}{{\sqrt {{\alpha _t}} }}\left( {{h_t} - \frac{{{\beta _t}}}{{\sqrt {1 - {{\bar \alpha }_t}} }}{\varsigma _\theta }\left( {{h_t},t} \right)} \right) ,
\end{equation}
where ${\varsigma _\theta }$ is the predicted noise by the neural network, which is used to recover ${\varsigma}$ from $h_t$, ensuring that ${\mu _\theta }$ closely approximates ${\tilde \mu _t}$. Accordingly, the network serves as a denoising model. It takes the noisy input $h_t$ and time step $t$ as inputs to predict the corresponding noise ${\varsigma _t}$. Consequently, the loss function at each time step is formulated as

\begin{equation}
\label{p14}
    {\mathcal{L}_{DM}}\left( \theta  \right) = {{\rm \mathbb{E}}_{t,{\varsigma _t}}}\left[ {\left\| {{\varsigma _t} - {\varsigma _\theta }\left( {{h_t},t} \right)} \right\|_2^2} \right] ,
\end{equation}
where the noisy input is constructed under the full masked diffusion paradigm ${\tilde h_t}$, and the corresponding formulation becomes~\eqref{p15}. After training, the denoising process can be performed iteratively using~\eqref{p16} to progressively remove noises, thus reconstructing the complete Alice–Jack paired fingerprints, where ${z_t} \sim \mathcal{N}\left( {0,\textbf{\textit{I}}} \right)$. The detailed steps of the CCMDM-based scheme are presented in Algorithm \ref{algorithm2}.

% \begin{equation}
% \label{p15}
%     {L_{DM}}\left( \theta  \right) = {{\rm E}_{{h_0},t,{\varsigma _t}}}\left[ {\left\| {m \cdot \left( {{\varsigma _t} - {\varsigma _\theta }\left( {\left( {1 - \left( {1 - \sqrt {{{\bar \alpha }_t}} } \right)m} \right) \cdot {{\tilde h}_0} + m \cdot \sqrt {1 - {{\bar \alpha }_t}} {\varsigma _t},t} \right)} \right)} \right\|_2^2} \right]
% \end{equation}

% \begin{equation}
% \label{p16}
%     {\tilde h_{t - 1}} = \left( {1 - m} \right) \cdot {\tilde h_t} + m \cdot \left[ {\frac{1}{{\sqrt {{\alpha _t}} }}\left( {{{\tilde h}_t} - \frac{{{\beta _t}}}{{\sqrt {1 - {{\bar \alpha }_t}} }}{\varsigma _\theta }\left( {m \cdot {{\tilde h}_t},t} \right)} \right) + {{\tilde \beta }_t}{z_t}} \right]
% \end{equation}

\subsection{Proposed CADM-based Fingerprint Generation Scheme}

\begin{rem}
\upshape
Although the previous CCMDM incorporates Jack's fingerprints $\hat{h}_J$ as concatenated inputs, it does not explicitly model the statistical dependency between Alice's fingerprints $h_0$ and Jack's fingerprints $\hat{h}_J$. As a result, CCMDM may not fully exploit the mapping information provided by $\hat{h}_J$, leading to limited sensitivity to the conditioning input and reduced accuracy in structural alignment and feature reconstruction of $h_0$. CCMDM essentially learns the form of conditional posterior as \( p(h_0 \mid h_T) = p(h_0, h_T) / p(h_T) \).
Therefore, even if the joint distribution $p\left( {{h_0},{h_T}} \right)$ or marginal distribution $p\left( {{h_T}} \right)$ is known during the sampling process, the conditional distribution remains uncertain.
% \begin{equation}
% \label{p18}
%     p\left( {{h_0}{{\left| h \right.}_T}} \right) = \frac{{p\left( {{h_0},{h_T}} \right)}}{{p\left( {{h_T}} \right)}}.
% \end{equation}
Moreover, since ${p_\theta }\left( {{h_{t - 1}}\left| {{h_t}} \right.} \right)$ is a probability distribution with variance in the reverse process, and each denoising step involves sampling a random variable ${h_{t - 1}}$ from this distribution as the input for the next step. Specifically,~\eqref{p16} reveals that the generated state ${h_{t - 1}}$ at each step consists of two components: a deterministic function $f\left( {{h_t},{\varsigma _\theta }\left( {{h_t},t} \right)} \right)$ predicted by the network and a stochastic perturbation term ${\tilde \beta _t}{z_t}$. The accumulation of these stochastic perturbations over the reverse process leads to a significant discrepancy between the generated Alice's fingerprint ${\tilde h_A}$ and the real fingerprint $h_0$. As a result, we develop CADM, which can fully use the mapping relationship between Alice's and Jack's CSI fingerprints to reduce the randomness during denoising.
\end{rem}

To enable accurate generation of Alice’s fingerprints ${\tilde h_A}$, in CADM, we incorporate Jack's fingerprints $\hat{h}_J$ as conditioning input, serving as characteristic-aligned conditional information rather than merely concatenated contextual input. This prior serves to guide the generation process toward realistic and structure-consistent CSI fingerprint outputs for Alice. Specifically, according to channel
extrapolation \cite{r34}, there exists an underlying functional mapping $\varphi:{\hat{h}_J} \to {h_0}$ between the Alice's fingerprints and Jack's fingerprints. Thus, for each specific $\hat{h}_J$, the distribution of Alice’s fingerprint $h_0$ is expected to concentrate around $\varphi\left( {{\hat{h}_J}} \right)$.

In the characteristic-aligned conditional denoising diffusion model, the predicted denoising mean at each step is more biased toward matching the distribution indicated by the condition $\varphi\left( {{\hat{h}_J}} \right)$, compared to the mean predicted ${\mu _\theta }\left( {h_t}\| {{\hat{h}_J}},t \right)$ by the standard masked diffusion approach.
Accordingly, based on \cite{zhou2025generative}, the mean in each step of the Gaussian reverse process can be reformulated as
\setcounter{equation}{27}

\begin{equation}
\label{p19}
    {\mu _\theta }\left( {{h_t},t,{\hat{h}_J}} \right) = \frac{1}{{\sqrt {{\alpha _t}} }}\left( {{h_t} - \frac{{{\beta _t}}}{{\sqrt {1 - {{\bar \alpha }_t}} }}{\varsigma _\theta }\left( {{h_t},t,{\hat{h}_J}} \right)} \right),
\end{equation}

\begin{algorithm}[t]
\caption{The Steps of the Proposed CADM-based Fingerprint Generation Scheme}
\label{algorithm3}
\begin{algorithmic}[1]
\State \textbf{Training Phase}
\State Construct 2-channel images for Alice $h_0$ and Jack $\hat{h}_J$, forming a paired input.
\Repeat{}
\State Sample Alice’s CSI fingerprints: ${h_0} \sim q(h_0)$
\State Sample diffusion step: $t \sim \text{Uniform}(1,\dots,T)$
\State Sample noise: $\varsigma \sim \mathcal{N}(0,\textit{\textbf{I}})$
\State Construct noisy sample: $h_t = \sqrt{\bar{\alpha}_t} h_0 + \sqrt{1 - \bar{\alpha}_t} \varsigma$
\State Extract multi-scale features from Jack's CSI fingerprint: $\{K_l, V_l\} = \text{Encoder}(\hat{h}_J)$
\State $\varsigma_\theta(h_t, t, \hat{h}_J) = \text{CrossAttn-UNet}(h_t, t, \{K_l, V_l\})$
\State Compute MSE loss: $\mathcal{L} = \left\| \varsigma - \varsigma_\theta(h_t, t, \hat{h}_J) \right\|^2$
\Until{convergence of $\mathcal{L}_{\text{cross-att}}$}
\vspace{1mm}
\State \textbf{Fingerprint Generation Phase}
\State Input: $h_T \sim \mathcal{N}(0,\textbf{\textit{I}})$, and conditioning Jack's CSI fingerprint $\hat{h}_J$
\For{$t = T,\dots,1$}
    \State Sample perturbation noise $z_t \sim \mathcal{N}(0,I)$ (only if $t > 1$)
    \State Extract $\{K_l, V_l\}$ from Jack's CSI fingerprint: $\{K_l, V_l\} = \text{Encoder}(\hat{h}_J)$
    \State Denoise via~\eqref{p22}
\EndFor
\State Return $\tilde{h}_A$ as the generated CSI fingerprint of Alice, conditioned on $\hat{h}_J$
\end{algorithmic}
\end{algorithm}

\noindent where ${\varsigma _\theta }\left( {{h_t},t,{\hat{h}_J}} \right)$ denotes the predicted noise residual. When the prior condition $\hat{h}_J$ is provided, the model tends to eliminate the noise components that cause the sample to deviate from the target distribution $\varphi\left( {{\hat{h}_J}} \right)$ during both the training and generation stages. As a result, the remaining signal is encouraged to align with the target ${\tilde h_A}$, as inferred from $\varphi:{\hat{h}_J} \to {h_0}$ \cite{r34}.

% \begin{figure*}[!b]
%     \hrulefill
%     \normalsize
%     \setcounter{equation}{31}
%     \begin{equation}
%     \label{O_encoder1}
%         \mathcal{O}\left(\sum_{l=1}^{L}\begin{pmatrix}n_r\begin{pmatrix}BC_lS_l\begin{pmatrix}C_{l-1}+C_l\end{pmatrix}+\delta_lB{S_l}^2C_l\end{pmatrix}\end{pmatrix}+\sum_{l=1}^{L-1}\begin{pmatrix}9B{C_l}^2S_{l+1}\end{pmatrix} \right)
%     \end{equation}
%     \setcounter{equation}{32}
%     \begin{equation}
%     \label{O_decoder1}
%         \mathcal{O}\left(\sum_{l=1}^{L}(n_r+1)\left(
%         B\, C_{L-l}\, S_{L-l} \left(C_{L-l+1} + C_{L-l}\right)+ \delta_l\, B\, S_{L-l}^2\, C_{L-l}\right)+ \sum_{l=2}^{L}9\, B\, C_{L-l}^2\, S_{L-l}\right)
%     \end{equation}
%     \setcounter{equation}{33}
% \end{figure*}

% \begin{figure*}[!b]
%     \hrulefill
%     \normalsize
%     \setcounter{equation}{34}
%     \begin{equation}
%     \label{O_encoder2}
%         \mathcal{O}\left(
%         \sum_{l=1}^{L} 2 n_r \left[ B C_l S_l (C_{l-1} + C_l) + B S_l^2 C_l \right]
%         + \sum_{l=1}^{L-1} 2 \cdot 9 B C_l^2 S_{l+1}
%         \right)
%     \end{equation}
%     \setcounter{equation}{35}
%     \begin{equation}
%     \label{O_decoder2}
%         \mathcal{O}\left(
%         \sum_{l=1}^{L} 2(n_r+1) \left[
%         B C_{L-l} S_{L-l} (C_{L-l+1} + C_{L-l}) + B S_{L-l}^2 C_{L-l}
%         \right]
%         + \sum_{l=2}^{L} 2 \cdot 9 B C_{L-l}^2 S_{L-l}
%         \right)
%     \end{equation}
% \end{figure*}
\begin{figure*}[!b]
    \hrulefill
    \normalsize
    \setcounter{equation}{31}
    % CCMDM Encoder
    \begin{equation}
    \label{O_encoder1}
        \mathcal{O}\left(\sum_{l=1}^{L}
        n_r\left[
            B C_l S_l (C_{l-1}+C_l)
            + \delta_l B S_l^2 C_l
        \right]
        +\sum_{l=1}^{L-1} 9 B C_l^2 S_{l+1} \right)
    \end{equation}
    % CCMDM Decoder
    \setcounter{equation}{32}
    \begin{equation}
    \label{O_decoder1}
        \mathcal{O}\left(
        \sum_{l=1}^{L}(n_r+1)\left(
            B\, C_{L-l}\, S_{L-l} (C_{L-l+1} + C_{L-l})
            + \delta_l B S_{L-l}^2 C_{L-l}
        \right)
        + \sum_{l=2}^{L} 9 B C_{L-l}^2 S_{L-l}\right)
    \end{equation}
    \setcounter{equation}{33}
\end{figure*}

\begin{figure*}[!b]
    \hrulefill
    \normalsize
    \setcounter{equation}{34}
    % CADM Encoder
    \begin{equation}
    \label{O_encoder2}
        \mathcal{O}\left(
        \sum_{l=1}^{L} n_r \Big[
            2 B C_l \tilde{S}_l (C_{l-1} + C_l)
            + 2\delta_l B \tilde{S}_l^2 C_l
            + 2\zeta_l B \tilde{S}_l^2 C_l
        \Big]
        + \sum_{l=1}^{L-1} 2 \cdot 9 B C_l^2 \tilde{S}_{l+1}
        \right)
    \end{equation}
    \setcounter{equation}{35}
    % CADM Decoder
    \begin{equation}
    \label{O_decoder2}
        \mathcal{O}\left(
        \sum_{l=1}^{L} 2(n_r+1) \left[
            B C_{L-l} \tilde{S}_{L-l} (C_{L-l+1} + C_{L-l})
            + \delta_l B \tilde{S}_{L-l}^2 C_{L-l}
            + \zeta_l B \tilde{S}_{L-l}^2 C_{L-l}
        \right]
        + \sum_{l=2}^{L} 2 \cdot 9 B C_{L-l}^2 \tilde{S}_{L-l}
        \right)
    \end{equation}
\end{figure*}

% \begin{figure*}[t]
%     \centering
%     \includegraphics[width=0.91\textwidth,trim=1 1 1 1,clip]{network_system.pdf}
%     \caption{U-Net architecture of the proposed CADM, with $ h_{A,t}$ as Alice’s noisy fingerprint and $h_J$ as Jack’s corresponding fingerprint. Cross-attention mechanism is applied for feature alignment in the encoder and decoder stages.}
%     \label{fig:ssim_auth}
% \end{figure*}

To further enable effective prior fusion, we introduce the cross-attention mechanism, originally proposed in the Transformer architecture \cite{vaswani2017attention}, into CADM to capture the dependencies between the collaborator's and the user's features. To balance model expressiveness and computational efficiency, each cross-attention layer employs 4 attention heads, which are empirically determined to provide sufficient representation capacity without incurring excessive computational cost. The detailed procedure is outlined in Algorithm \ref{algorithm3}. After converting the CSI fingerprint data into image-like representations, a conditional U-Net \cite{ronneberger2015u}, which is characterized by a symmetric encoder-decoder architecture with skip connections for capturing multi-scale spatial features, within the diffusion model is employed to guide the noisy CSI fingerprint of Alice and the fingerprint of Jack for structurally aligned feature mapping. The network consists of a dual-branch encoder-decoder architecture. Each branch contains an encoder with 17 two-dimensional convolutional layers and 3 downsampling operations, enabling both the noisy Alice channel $h_t$ and the Jack channel $\hat{h}_J$ to be encoded with the same structure for dimensionality reduction and feature extraction. Specifically, the current noisy diffusion state feature of Alice $h_t$, is used as the Query, while the conditional fingerprint $\hat{h}_J$ provided by Jack, serves as both the Key and Value. In the encoder, $\hat{h}_J$ is processed into multi-scale contextual features $\left\{ {{K_l},{V_l}} \right\}$ that are feature-wise and dimensionally aligned with Alice's intermediate feature representations $Q_l$ at various layers. In this way, the cross-attention module adaptively focuses on different structural and value components of $h_t$ based on $\hat{h}_J$, thus embedding Jack's channel information into the noise prediction process ${\varsigma _\theta }\left( {{h_t},t,{\hat{h}_J}} \right)$. The attention mechanism is incorporated according to the following formulation

\setcounter{equation}{28}
\begin{equation}
\label{p20}
    Attention\left( {Q,K,V} \right) = {\rm{softmax}}\left( {\frac{{Q{K^T}}}{{\sqrt {{d_k}} }}} \right)V .
\end{equation}

This method can actively query the structural features of the Jack region when denoising each spatial position, thereby learning an implicit mapping function ${h_0} = \varphi\left( {{\hat{h}_J}} \right)$ \cite{r34} between the two channels. Accordingly, the loss function of the cross-attention-based conditional generation is reformulated as

\begin{equation}
\label{p21}
    {\mathcal{L}_{cross - att}} = {\mathbb{E}_{{h_0},{h_J},{\varsigma _t},t}}\left[ {\left\| {{\varsigma _t} - {\varsigma _\theta }\left( {{h_t},t,{\hat{h}_J}} \right)} \right\|_2^2} \right] .
\end{equation}

Then, the same as the masked diffusion process, we can represent the denoising process one step by step as 

\begin{equation}
\label{p22}
    h_{t-1} = \frac{1}{\sqrt{\alpha_t}} \left( h_t - \frac{\beta_t}{\sqrt{1 - \bar{\alpha}_t}} \varsigma_\theta(h_t, t, \hat{h}_J) \right) + \tilde{\beta}_t z_t .
\end{equation}

\subsection{Complexity Analysis of the Proposed Schemes}

\subsubsection{Complexity Analysis of CCMDM}

Let the channel multiplication sequence be denoted by $\mathcal{P}=\{p_1, p_2, \ldots, p_L\}$, where $L$ is the number of layers. Denote $B$, $\gamma$, and $C_b$ as the batch size, time channel expansion factor, and base channel size, respectively. The input is formed by concatenating two CSI fingerprints along the last dimension, resulting in an input tensor of size $(C, H, 2W)$. Then, the complexity of the time-embedding layer is $\mathcal{O}\left( B \left(\gamma {C_b}^2 +\gamma^2{C_b}^2 \right)\right)$. Since the convolution operation adopts a kernel of $3 \times 3$, the complexity of the first convolution layer can be expressed as $\mathcal{O}\left(9B{C_b}C S\right)$, where $C$ denotes the number of input channels, and $S=H\times 2W$ is the spatial resolution. 
The encoder part includes $L$ layers. As a result, the complexity of the encoder process can be expressed in~\eqref{O_encoder1}, where $n_r$ is the number of ResNetBlocks in each layer, $C_l=C_b\cdot p_l$ is the input channel number of the $l^{th}$ layer, $C_0 = C$, and $S_l$ is the spatial resolution at the $l^{th}$ layer. $\delta_l=1$ represents that self-attention is employed at the $l^{th}$ layer. Similarly, the complexity of the decoder process is expressed in~\eqref{O_decoder1}. 
% Here, $C_{L-l}$ and $S_{L-l}$ denote the output channel number and spatial resolution at the $(L{-}l)$-th encoder layer, corresponding to the $l$-th decoder layer. $C_{L-l+1}$ is the output channel of the previous decoder layer, and the total input channel of each decoder block is $C_{L-l+1} + C_{L-l}$ due to skip connection concatenation. Other notations are consistent with the encoder analysis.
The complexity of the bottleneck is $\mathcal{O}\left(2B\, C_L^2\, S_L+ 2B\, \gamma\, C_b\, C_L+ B\, S_L^2\, C_L\right)$. To further provide an intuitive understanding of the computational scale, we aggregate the complexitiy of all modules and retain only the dominant terms. Since the convolutional and attention operations are main contributors to the overall computational cost, the total complexity of CCMDM is approximated as

\setcounter{equation}{33}
\begin{equation}
    \label{O_CCMDM}
        \mathcal{O}_\text{CCMDM} = \mathcal{O}\left( T  B  L  n_r  \left( C_{\max}^2 S_{\max} +  S_{\max}^2 C_{\max} \right)\right),
\end{equation}
where $C_{\max}$ and $S_{\max}=H\times 2W$ denote the maximum number of channels and the maximum spatial resolution across all layers, respectively. $T$ is the total number of diffusion steps.

\subsubsection{Complexity Analysis of CADM}

In CADM, the input consists of two separate CSI fingerprints, $h_{A,t}$ and $h_J$, each of size $(C, H, W)$, which are processed in parallel by dual encoder-decoder branches. At every ResNet block in both the encoder and decoder, a cross-attention module is added between the two branches, while self-attention is retained as in CCMDM. Thus, the complexity at each block includes both self-attention and cross-attention terms, and all operations are performed on feature maps of size $H \times W$ for each branch.
The complexity of the time-embedding and first convolutional layers doubles due to the dual-branch design: $\mathcal{O}\left(2 B (\gamma {C_b}^2 + \gamma^2 {C_b}^2) \right)$ and $\mathcal{O}\left(2 \cdot 9 B C C_b \tilde{S}\right)$, where $\tilde{S} = H \times W$.
The encoder complexity is given in~\eqref{O_encoder2}, where, for each block, the self-attention term $\delta_l$ and cross-attention term $\zeta_l$ (typically $\zeta_l=1$ for all blocks) are both included.
The bottleneck contains two ResNetBlocks per branch and a cross-attention operation, leading to the complexity $\mathcal{O}\left(4B\, C_L^2\, \tilde{S}_L + 4B\, \gamma\, C_b\, C_L + 2B\, \tilde{S}_L^2\, C_L \right)$.
Similarly, the decoder complexity is expressed in~\eqref{O_decoder2}.
Aggregating all components and retaining only the dominant terms, the total computational complexity of CADM is given by

\setcounter{equation}{36}
\begin{equation}
    \label{O_CADM}
        \mathcal{O}_\text{CADM} = \mathcal{O}\left( T  B  L  n_r  \left( 2C_{\max}^2 \tilde{S}_{\max} + 4\tilde{S}_{\max}^2 C_{\max} \right)\right),
\end{equation}
where $C_{\max}$ and $\tilde{S}_{\max} = H \times W$ denote the maximum channel number and spatial resolution in each branch, respectively. The factor $2$ reflects the dual-branch design, while the factor $4$ in the attention term arises from the addition of cross-attention alongside self-attention in every block.

% \begin{rem}
% \upshape
% Although the main complexity terms for the concatenated single-branch (CCMDM) and dual-branch (CADM) architectures may appear similar when accounting for the effective spatial resolution and total number of channels, in practice, the dual-branch model incurs significantly higher parameter storage, memory consumption, and hardware parallelization overhead. The CADM design maintains two independent parameter sets and performs parallel computations for each input branch, as well as additional cross-attention modules, which cannot be strictly equated to a single-branch network with a wider input. 
% Experimental results in the simulation section also show that the training speed and GPU resource usage of CADM are consistently lower and higher, respectively, than those of CCMDM, even for similar theoretical FLOPs.    
% \end{rem}
% Compared to CCMDM, the CADM scheme incurs approximately twice the computational cost due to its dual-branch and cross-attention design.

\subsubsection{Latency and Power Consumption Analysis of CADM}

To assess the feasibility of the proposed scheme in practical systems, we evaluated the inference latency and energy consumption on a high-performance workstation configured with an NVIDIA RTX A6000 GPU and an Intel Xeon w7-2495X CPU.

\begin{enumerate} % 需要 \usepackage{enumitem}，显示为 (1), (2)
    % 第一项
    \item[(a)] \textbf{Inference Latency:} The measured average inference latency is approximately 9.03 ms per time step f denoising. While standard diffusion requires one thousand steps, by adopting the Denoising Diffusion Implicit Models (DDIM) accelerator with 20 sampling steps, the total E2E inference time is reduced to approximately 180.6 ms. This falls well within the sub-second authentication latency requirements for many 6G applications.
    
    % 第二项 (您的代码里有个 \item 但没内容，根据上下文应该是 Energy Consumption)
    \item[(b)] \textbf{Energy Consumption:} Power Consumption Estimation: Based on the hardware specifications, the Thermal Design Power is 300 W for the NVIDIA RTX A6000 and 225 W for the Intel Xeon w7-2495X, resulting in a total peak system power of approximately 525 W. Consequently, the estimated energy consumption for a single authentication instance (180.6 ms) is approximately 94.8 J, which is a conservative upper-bound estimate based on a general-purpose workstation. In actual 6G deployment, utilizing dedicated edge AI chips (e.g., FPGA or ASIC) would significantly further reduce both latency and power consumption.
    
\end{enumerate}

\section{Simulation Results and Analysis}

\subsection{Simulation Settings}

\begin{table}[tbp]\footnotesize
    \caption{DeepMIMO `O1' Scenario Parameters \cite{Alkhateeb2019} 
    \label{tab:table1}}
    \centering
   \begin{tabular}{ll}
        \toprule
        Parameter & Value\\
        \midrule
        Receiver antenna array size & $1 \times 2 \times 4$ \\
        Transmitter antenna array size & $1 \times 2 \times 4$ \\
        Antenna element spacing & $0.5\lambda$ \\
        Receiver antenna angle (x, y, z) & (5$^\circ$, 10$^\circ$, 20$^\circ$) \\
        Transmitter antenna angle (x, y, z) & (0$^\circ$, 30$^\circ$, 0$^\circ$) \\
        Radiation pattern & Isotropic \\
        Bandwidth & 50 MHz \\
        Frequency & 28 GHz \\
        Number of OFDM subcarriers & 32 \\
        Number of paths & 5 \\
        OFDM sampling factor & 1 \\
        OFDM subcarrier limit & 32 \\
        \bottomrule
    \end{tabular}
\end{table} 

\begin{figure}[tbp]
    \centering
    \includegraphics[width=0.4\textwidth,trim=15 15 15 15,clip]{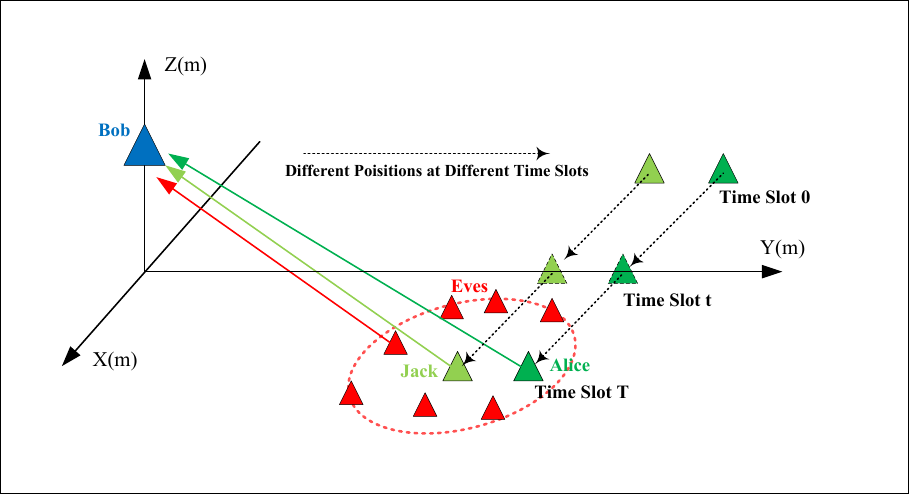}
    \caption{The positions of nodes involved, where Alice, Bob, and Jack are posited at (23.5 m, 266.93 m, 0), (0, 0, 4 m), and (23.5 m, 266.73 m, 0), respectively. Eves are randomly distributed within a circle centered at Alice with radius $r$.}
    \label{system_model}
\end{figure}

To emulate practical dynamic communication scenarios, we adopt dynamic outdoor channel parameters from the 'O1' scenario of the DeepMIMO dataset \cite{Alkhateeb2019} to generate time-varying channel responses of Alice and Jack. Specifically, to construct the dataset, we extracted a total of 12000 CSI samples based on the continuous movement trajectories of Alice and Jack within the dataset grid. The data samples were then partitioned into a training set and a testing set with a ratio of 9:1. In addition, to emulate environmental variations in dynamic scenarios, Alice and Jack are placed at different locations at each time step and move at constant velocities to ensure channel continuity. The key scenario parameters of DeepMIMO are summarized in Table \ref{tab:table1}. The positions of involved nodes are illustrated in Figure \ref{system_model}, where Alice and Jack move at a constant speed in the same direction while maintaining a fixed relative distance. An adversary, Eve, is randomly placed within a circular area centered on Alice with radius $r$. To evaluate the authentication performance against frequent spoofing, we initialized 5 candidate Eves randomly distributed in this area. During the evaluation, the attack frequency ratio $k$ (as defined in \eqref{pla_k}) was set to 0.5, where the number of attack samples equals the number of legitimate samples. At each attack instance, one Eve is randomly selected from the candidates to transmit spoofing signals, thereby simulating a diverse and high-intensity attack environment.

\begin{figure}[tbp]
    \centering
    \includegraphics[width=0.4\textwidth,trim=1.5 1.5 1.5 1.5,clip]{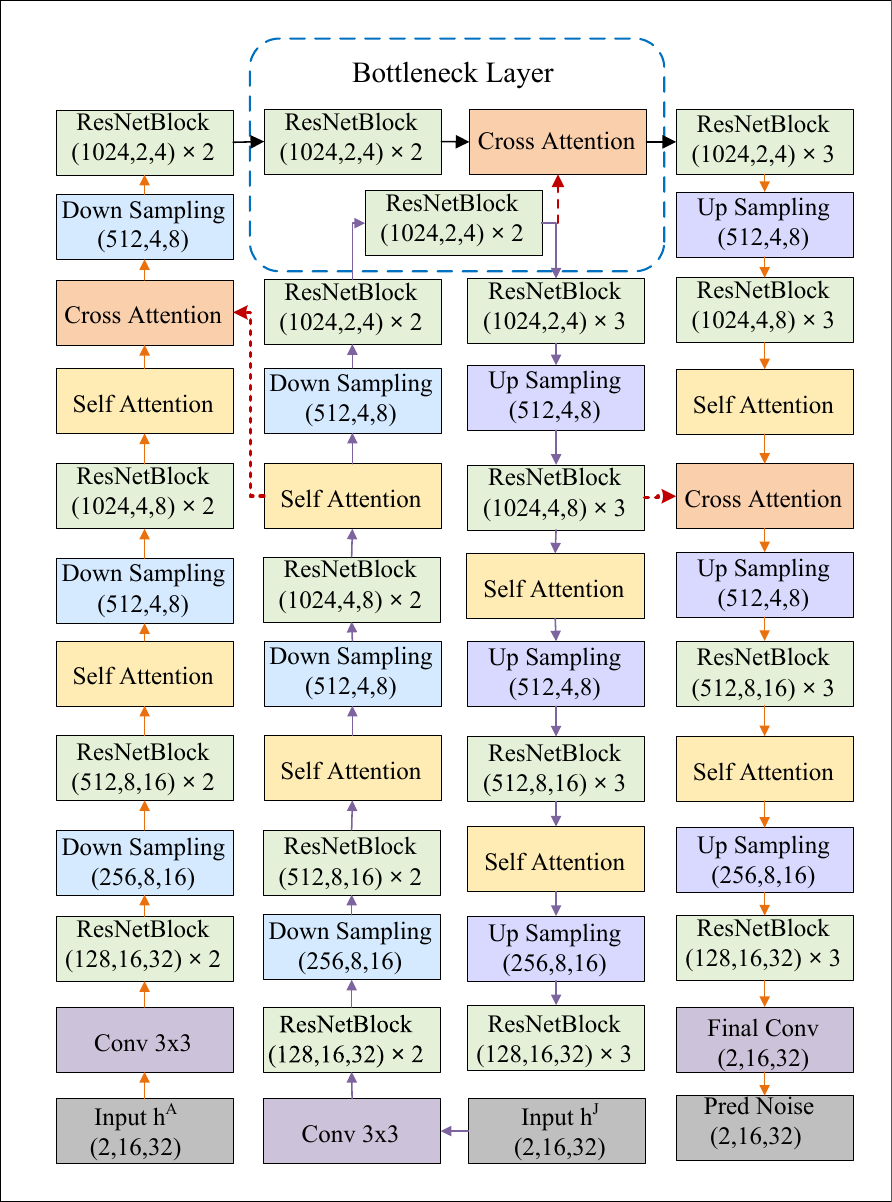}
    \caption{Illustration of the proposed CADM scheme, where the U-Net framework consists of an encoder and a decoder, each comprising three downsampling and three upsampling modules. The network integrates cross-attention and self-attention modules to fuse features 
    across multiple scales for enhanced CSI fingerprint generation.}
    \label{network_simulation}
\end{figure}

\begin{table}[tbp]\footnotesize
    \caption{Model Hyperparameter Settings \label{tab:table2}}
    \centering
   \begin{tabular}{ll}
        \toprule
        Parameter & Value\\
        \midrule
        Diffusion timesteps & 1000 \\
        Input shape & $(2, 16, 32)$ \\
        Training epochs & 1600 \\
        Batch size & 128 \\
        Learning rate & $1\times10^{-4}$ \\
        Base channel size & 64 \\
        Channel multiplier & (1, 2, 4, 4) \\
        Self-Attention  & (True, False, False, True) \\
        Num\_heads of Cross-Attention & 4 \\
        Dropout rate & 0.1 \\
        Time embedding multiplier & 4 \\
        \bottomrule
    \end{tabular}
\end{table}

% \begin{figure}[t]
%     \centering
%     \includegraphics[width=0.5\textwidth,trim=15 15 15 15,clip]{system_model.pdf}
%     \caption{Simulation illustration of identity authentication for a moving Alice assisted by a neighboring Jack. The figure shows the positions of Alice (23.5, 266.93, 0), Bob (0, 0, 4), and Jack (23.5, 266.73, 0), with Eves randomly distributed within a circle centered at Alice with radius $r$.}
%     \label{fig2:"Simulation illustration of identity authentication of dynamic Alice based on neighboring Jack"}
% \end{figure}

% Figure \ref{network_simulation} illustrates the overall architecture of the proposed CADM, and Table \ref{tab:table2} illustrates the hyperparameter of CADM. The network is built upon a U-Net framework and consists of three main components: an encoder, a bottleneck, and a decoder. Multiple ResNet modules and attention mechanisms are embedded within both the encoding and decoding paths to enhance feature extraction and reconstruction capabilities. In particular, a cross-attention mechanism is introduced to model the dependency between the channel features of Alice and Jack, enabling feature alignment both in multi-scale fusion and at the bottleneck layer. The network takes the noisy input channel $\tilde h^A_t$ as a condition and outputs the predicted noise component $\hat n^A_t$, which is subsequently removed to obtain the denoised channel.
Figure \ref{network_simulation} illustrates the overall architecture of the proposed CADM, and Table \ref{tab:table2} details its hyperparameters. The network is built upon a U-Net framework and consists of three main components: an encoder, a bottleneck, and a decoder. Multiple ResNet modules and attention mechanisms are embedded within both the encoding and decoding paths to enhance feature extraction and reconstruction capabilities. The network takes the noisy input channel $h_t$ and the collaborator's channel $\hat{h}_J$ as inputs and outputs the predicted noise component $\hat n^A_t$, which is subsequently removed to obtain the denoised channel.
In essence, this architecture in Figure \ref{network_simulation} is specifically tailored to realize the physical concept of spatial-domain channel extrapolation through three integrated design strategies. First, a dual-branch encoder structure is employed to explicitly model the collaborator's channel $\hat{h}_J$ as a conditional input, structurally enforcing the learning of the spatial mapping from Jack to Alice. Second, the Cross-Attention mechanism serves as the core algorithmic innovation; it uses Alice's noisy features as queries to adaptively align with the collaborator's features, effectively overcoming signal misalignment between these spatially correlated channels. Finally, by embedding these mechanisms across multiple resolution levels, the model fuses auxiliary information at both semantic and fine-grained textural scales, ensuring high fidelity in reconstructing the complex multipath details essential for authentication.

% \begin{figure}[t]
%     \centering
%     \includegraphics[width=0.40\textwidth,trim=0 0 0 0,clip]{network_simulation.png}
%     \caption{Illustration of the proposed CADM scheme, where XX, XX, and XX. The network integrates cross-attention and self-attention modules to fuse features from Alice’s noisy fingerprint $\tilde h_{A,t}$ and Jack’s fingerprint $h_J$ across multiple scales for enhanced channel generation.}
%     \label{fig2:"Simulation illustration of identity authentication of dynamic Alice based on neighboring Jack"}
% \end{figure}

\subsection{Comparative Schemes}

% To comprehensively evaluate the performance of the proposed user-domain channel extrapolation-based generation and physical-layer authentication method, four representative baseline approaches are selected for comparison, covering both time domain modeling schemes and AI model design perspectives.

\begin{itemize}
\item \textbf{GRU-based PLA \cite{R36}:}
GRU introduces update and reset gates to capture critical temporal sequence information.
\item 
\textbf{LSTM-based PLA \cite{r35}:}
Compared to GRU, LSTM is better at modeling long-term dependencies.
\item 
\textbf{The Scheme Using Collaborator's Fingerprints as Alice's Fingerprints (CA):}
This scheme does not leverage GAI to predicted Alice's fingerprints, but directly uses Jack's CSI fingerprints as Alice's fingerprints. 
\item 
\textbf{CAVAE-based Fingerprint Generation:}
This scheme integrates VAE with a cross-attention mechanism.

\end{itemize}
% \subsubsection{GRU-based PLA \cite{R36}}
% GRU introduces update and reset gates to capture and retain critical information in temporal sequences. Therefore, GRU utilizes the continuously time-varying nature of Alice’s fingerprints during device movement to predict the fingerprint at the next time step, which is then used for identity authentication.
% \subsubsection{LSTM-based PLA \cite{r35}}
% Compared to GRU, LSTM introduces an independent memory cell along with three gating mechanisms: input gate, forget gate, and output gate, offering enhanced capability to model long-term dependencies. LSTM is used to predict Alice's CSI fingerprints. 
% \subsubsection{The Scheme Using Collaborator's Fingerprints as Alice's Fingerprints (CA)}
% Considering the spatial correlation between adjacent users in the user domain, the CA scheme directly uses Jack's CSI fingerprints as Alice's fingerprints. This scheme does not leverage GAI to predicted Alice's fingerprints.
% \subsubsection{Cross-Attention VAE-based Fingerprint Generation}
% Cross-Attention VAE (CAVAE) integrates VAE with a cross-attention mechanism.
% Jack's CSI fingerprints at the same time slot is used as a conditional input, aligned and fused with the latent representation of Alice's fingerprint to capture spatial correlation and structural dependency between them. In the decoding phase, the model reconstructs Alice’s channel using the combined latent and conditional features, enabling device authentication for Alice.

\subsection{Performance Metrics}

% To comprehensively evaluate the effectiveness of the proposed channel generation and physical-layer authentication scheme, five performance metrics are adopted in this work. These metrics assess the model from five dimensions including structural similarity, amplitude fidelity, spatial consistency, error measurement and authentication capability.

\begin{figure*}[tbp]
    \centering
    \includegraphics[width=0.9\textwidth,trim=15 15 15 15,clip]{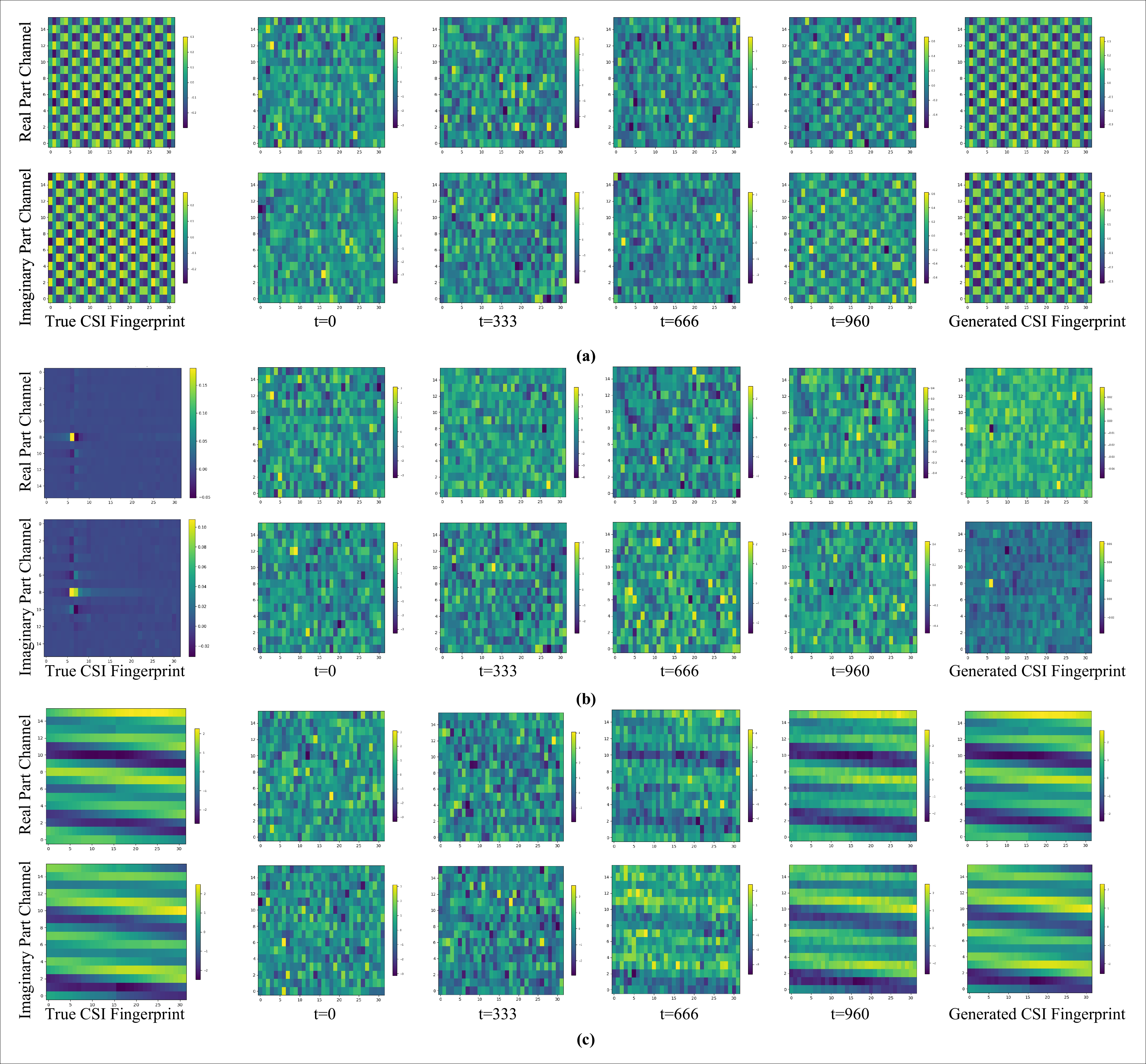}
    \caption{The visualization of the denoising process during CSI fingerprint generation. The upper layer of (a) (b) (c) is the real part channel of the generated fingerprint, and the layer below is the imaginary part. $t$ is the denoising time step. (a) is the DeepMIMO O1 outdoor frequency domain CSI fingerprint, (b) is the DeepMIMO O1 outdoor angle delay domain CSI fingerprint, (c) is the DeepMIMO I3 indoor frequency domain CSI fingerprint.}
    \label{DIff_process}
\end{figure*}

\begin{figure}[tbp]
    \centering
    \includegraphics[width=0.4\textwidth,trim=0 0 0 0,clip]{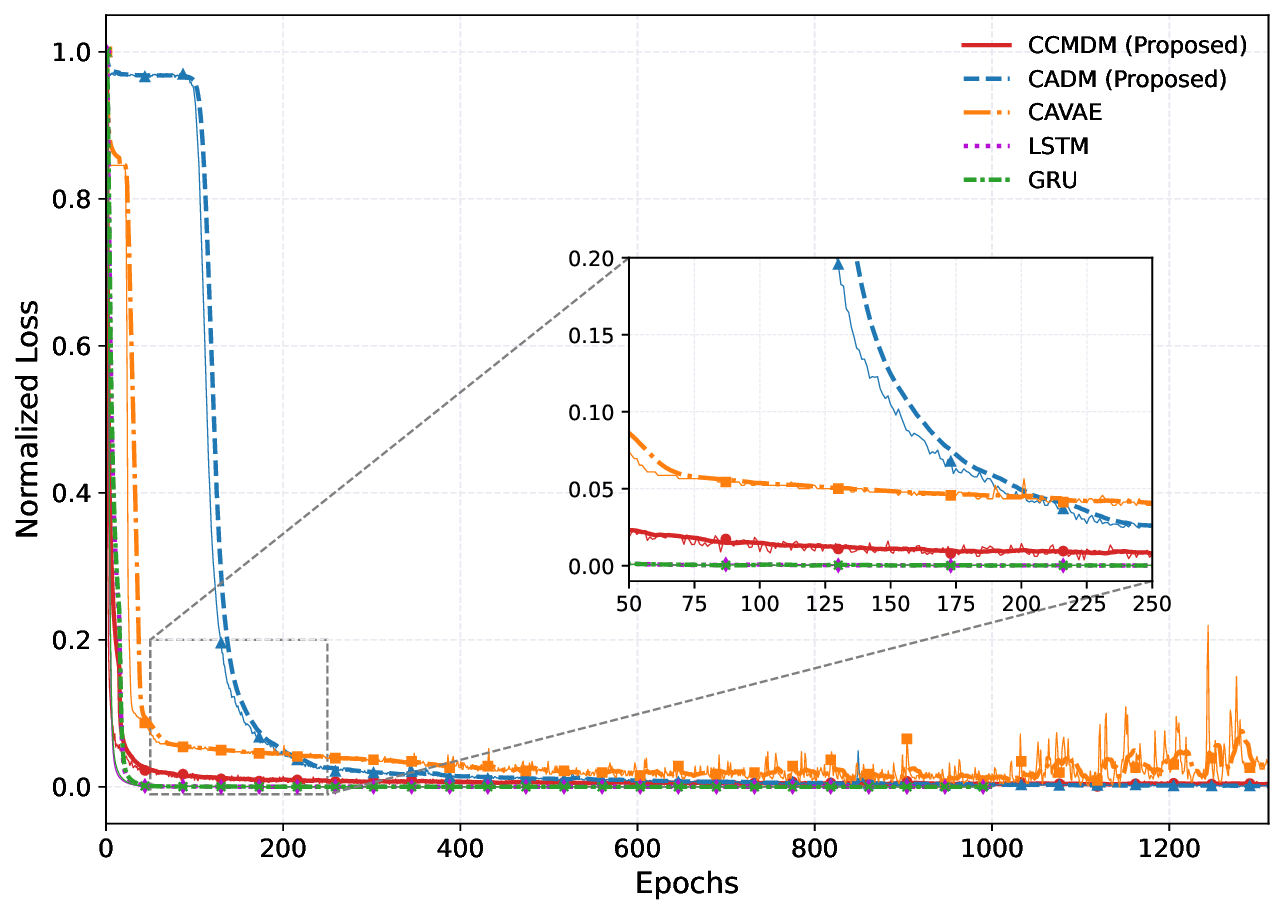}
    \caption{Comparison of the normalized loss convergence of different models. The average loss line is obtained using a moving average with a window size of 20.}
    \label{loss_curve_compare_smooth_fixedlength}
\end{figure}

\subsubsection{Performance Metrics for CSI Fingerprint Generation}

\begin{itemize}
\item \textbf{SSIM:}
SSIM measures the perceptual similarity between the generated channel image and the real channel image. It takes into account luminance, contrast, and structural information.
\item \textbf{PSNR:}
PSNR is used to measure the pixel-level amplitude difference between the generated image and the reference image. A higher PSNR value indicates smaller error and better image fidelity.
\item \textbf{Cosine Similarity:}
Cosine similarity measures the directional similarity between two vectors while ignoring differences in magnitude. It is suitable for comparing the structural distribution of CSI fingerprint images.
\item \textbf{NMSE:}
NMSE measures the ratio of the amplitude error between the generated CSI fingerprint and the real CSI fingerprint to the energy of the real. A smaller value indicates higher prediction accuracy. NMSE is more sensitive to absolute value differences and better reflects the accuracy of amplitude reconstruction.
\end{itemize}

\subsubsection{Performance Metrics for PLA}
\begin{itemize}
\item \textbf{F1 Score:}
Motivated by \cite{meng2023physical} and \cite{xia2022multiple}, F1 score is used to comprehensively evaluate the authentication performance. 
It is defined as

\begin{equation}
\label{F1}
    \rm{F1} = \frac{{2 \cdot {\rm{Precision}} \cdot {\rm{Recall}}}}{{{\rm{Precision}} + {\rm{Recall}}}},
\end{equation}
where
$\rm{Precision} = \frac{\rm{TP}}{\rm{TP + FP}}$
and 
$\rm{Recall} = \frac{\rm{TP}}{\rm{TP + FN}}$.
TP (True Positive) refers to the number of samples correctly identified as Alice. FP (False Positive) refers to non-Alice samples incorrectly identified as Alice. FN (False Negative) refers to real Alice samples that were not correctly identified.

\item \textbf{Authentication Error Rate $R_E$:}
$R_e\in[0,1]$~\cite{meng2025efficient} is the ratio of fingerprint samples with incorrect authentication to all fingerprint samples.
\end{itemize}
% Motivated by \cite{meng2023physical} and \cite{xia2022multiple}, F1 score is used to comprehensively evaluate the authentication performance. 
% It is defined as

% \begin{equation}
% \label{F1}
%     \rm{F1} = \frac{{2 \cdot {\rm{Precision}} \cdot {\rm{Recall}}}}{{{\rm{Precision}} + {\rm{Recall}}}},
% \end{equation}
% where
% $\rm{Precision} = \frac{\rm{TP}}{\rm{TP + FP}}$
% and 
% $\rm{Recall} = \frac{\rm{TP}}{\rm{TP + FN}}$.
% TP (True Positive) refers to the number of samples correctly identified as Alice. FP (False Positive) refers to non-Alice samples incorrectly identified as Alice. FN (False Negative) refers to real Alice samples that were not correctly identified.

\subsection{Simulation Results of Fingerprint Prediction}

\subsubsection{Visualization of CSI Fingerprint Generation} Figure \ref{DIff_process} illustrates the reverse generation of Alice's CSI fingerprints using the CADM model at 20 dB SNR. As observed in the outdoor scenario frequency domain CSI fingerprints of Fig. \ref{DIff_process} (a) and the indoor scenario frequency domain CSI fingerprints of Fig. \ref{DIff_process} (c), the generated frequency-domain CSI fingerprints align closely with the ground truth in both structural distribution and amplitude range, confirming accurate reconstruction. In contrast, the angle-delay domain generated CSI fingerprints in Fig. \ref{DIff_process} (b) exhibit slight background noise. This stems from the inherent sparsity of the domain where signal energy concentrates in few components with significant magnitude gaps compared to the background. Nevertheless, the accurate recovery of key signal positions validates the model's effectiveness, while future log-domain pre-processing can effectively mitigate these background magnitude mismatches.

\subsubsection{Loss Convergence Analysis of Proposed and Comparative Schemes}
Figure \ref{loss_curve_compare_smooth_fixedlength} illustrates the loss convergence trends of the proposed CCMDM and CADM schemes alongside the comparative schemes (CAVAE, LSTM, and GRU) during training. Since these models employ distinct loss functions, such as noise prediction error for diffusion schemes and reconstruction error for LSTM, GRU, and CAVAE, their absolute loss values differ. Therefore, we normalized the loss curves of all models to the range of $[0, 1]$ to facilitate a fair comparison of convergence speed and stability.

It can be observed that the time-series models, LSTM and GRU, exhibit the most rapid convergence, which is attributed to their relatively simple architectures and direct optimization objectives. Among the generative models, the proposed CCMDM demonstrates a significant advantage in convergence speed, passing the rapid descent phase and stabilizing around 200 epochs. In contrast, CADM requires a longer warm-up period, and finally converges after approximately 1200 epochs. This disparity arises because, unlike the simple channel concatenation in CCMDM, the cross-attention mechanism in CADM necessitates optimizing additional projection matrices to learn soft feature alignment, which imposes a higher learning burden. However, as demonstrated in the subsequent generation quality analysis, this additional training cost of CADM translates into superior fingerprint generation quality. Furthermore, while CAVAE converges quickly in the initial stage, it exhibits noticeable fluctuations in the later stages, suggesting that its training stability is inferior to that of the diffusion-based CCMDM and CADM.

\begin{figure}[t]
    \centering
    \includegraphics[width=0.4\textwidth,trim=0 0 0 0,clip]{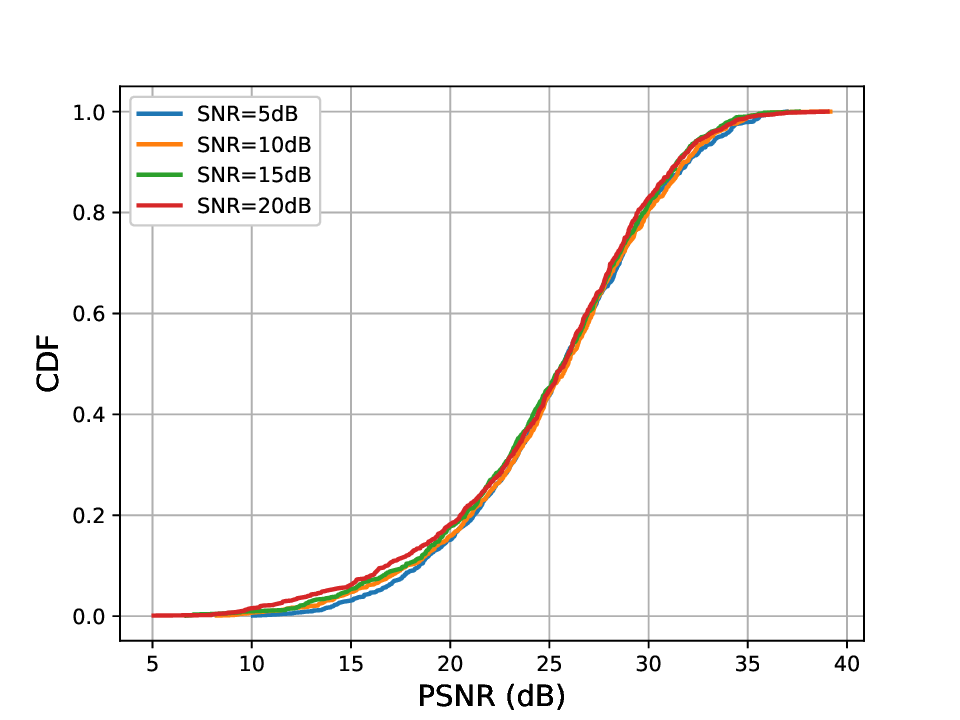}
    \caption{Fingerprint generation performance of the proposed CADM-based scheme in PSNR at SNR = 5 dB, 10 dB, 15 dB, and 20 dB.}
    \label{PSNR_CDF}
\end{figure}

\begin{figure}[t]
    \centering
    \includegraphics[width=0.4\textwidth,trim=0 0 0 0,clip]{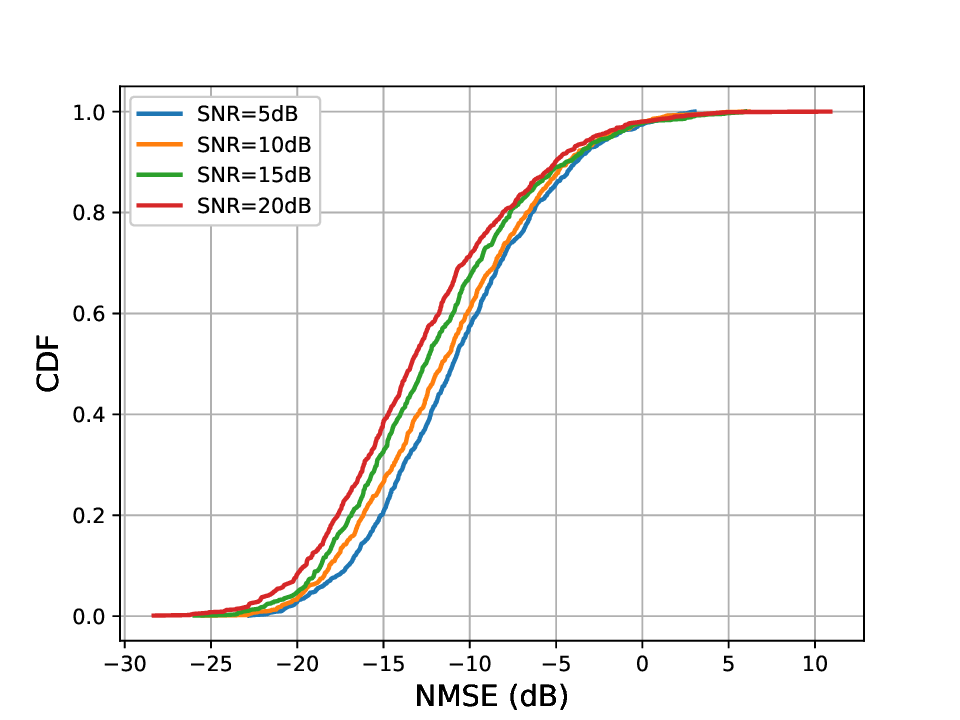}
    \caption{Fingerprint generation performance of the proposed CADM-based scheme in NMSE at SNR = 5 dB, 10 dB, 15 dB, and 20 dB.}
    \label{nmse_CDF}
\end{figure}

\begin{figure}[t]
    \centering
    \includegraphics[width=0.4\textwidth,trim=0 0 0 0,clip]{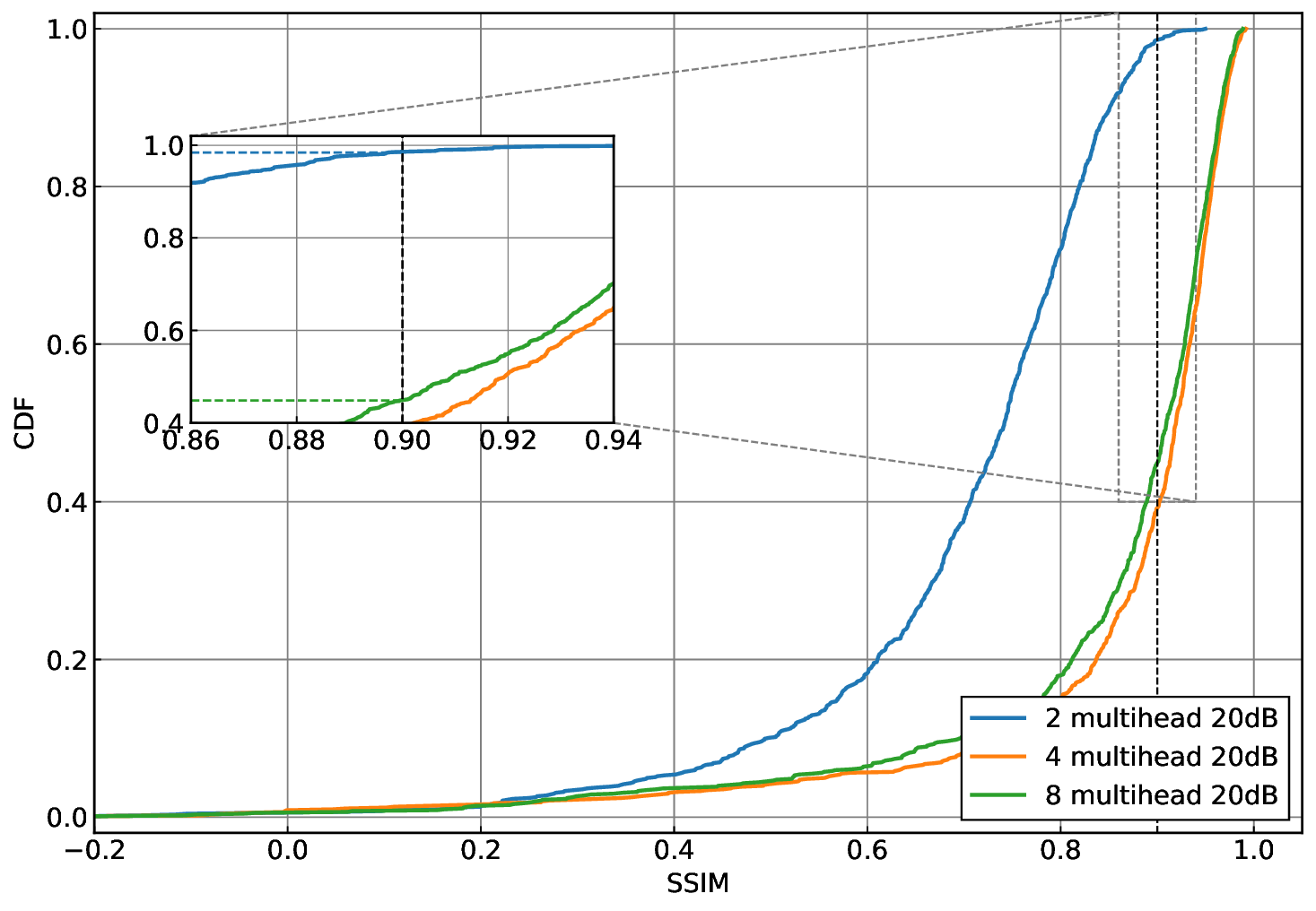}
    \caption{Fingerprint generation performance of the proposed CADM-based scheme at different cross-attention multiheads.}
    \label{cdf_multihead_ssim_final}
\end{figure}

\begin{figure}[t]
    \centering
    \includegraphics[width=0.4\textwidth,trim=0 0 0 0,clip]{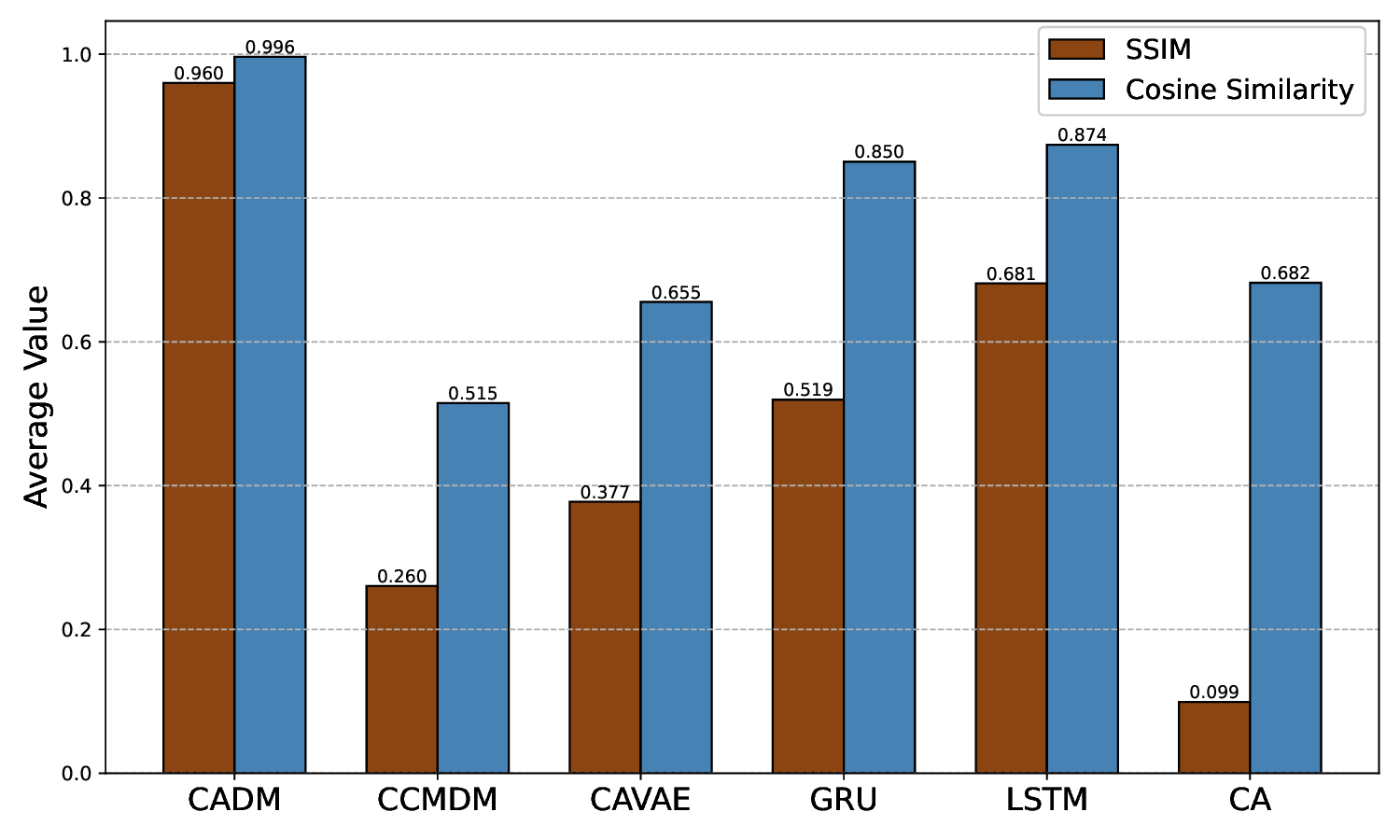}
    \caption{Fingerprint generation performance of the CCMDM-based and CADM-based schemes compared with comparative schemes in SSIM and Cosine Similarity when SNR = 20 dB.}
    \label{SSIM_Cosine}
\end{figure}

\begin{figure}[t]
    \centering
    \includegraphics[width=0.4\textwidth,trim=0 0 0 0,clip]{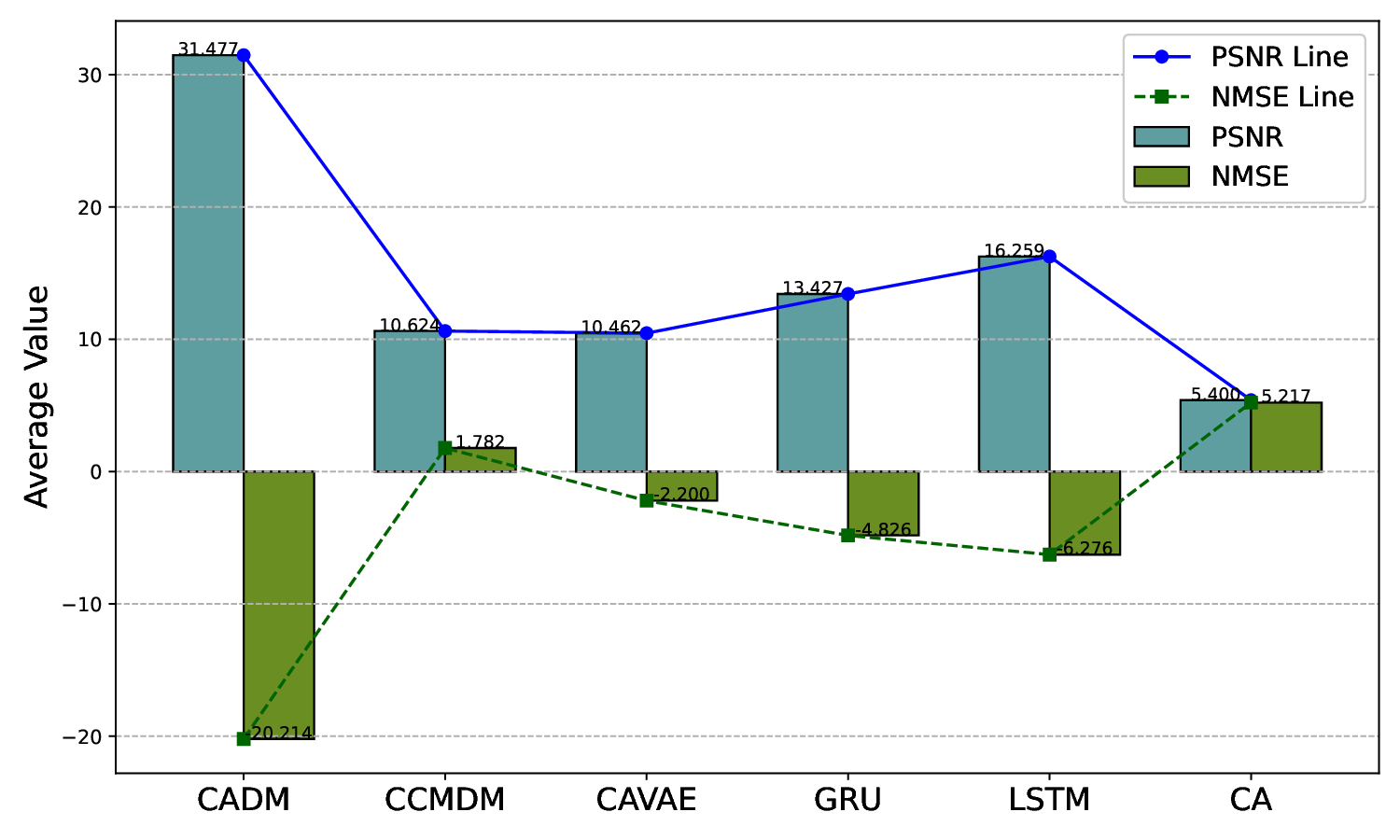}
    \caption{Fingerprint generation performance of the CCMDM-based and CADM-based schemes compared with comparative schemes in PSNR and NMSE when SNR = 20 dB.}
    \label{PSNR_NMSE}
\end{figure}

\subsubsection{Noise Robustness of the Proposed CADM-based Fingerprint Generation Scheme}
Figure \ref{PSNR_CDF} and Figure \ref{nmse_CDF} compare the CDF curves of PSNR and NMSE for the channel matrices generated by CADM under different SNR conditions. As shown in Figure \ref{PSNR_CDF}, the PSNR curves exhibit minimal variation with increasing SNR, and the distributions largely overlap across all SNR levels. This indicates that the proposed model consistently produces CSI fingerprint images with high structural fidelity across varying noise levels. Since PSNR primarily reflects overall amplitude consistency, it is less sensitive to small or localized distortions. As a result, PSNR may not adequately capture subtle degradations in generation quality under varying noise intensities.
In contrast, NMSE, as a relative error metric, is more sensitive to amplitude-level deviations. Therefore, Figure \ref{nmse_CDF} shows more pronounced differences in the NMSE curves. As SNR increases, the CDF curves shift to the left, indicating a clear reduction in reconstruction error. In particular, under high SNR conditions such as 20 dB, NMSE values concentrate in lower negative ranges, demonstrating improved precision in channel modeling. This is because NMSE is expressed in dB scale, where more negative values correspond to smaller normalized reconstruction errors and thus better modeling accuracy. Moreover, although the NMSE distribution under 5 dB SNR is not as favorable as that under 20 dB, the difference remains relatively small. For example, at $\rm{NMSE} = -15dB$, the gap between the two distributions is only about 0.1. This demonstrates that the proposed method maintains strong reconstruction capability even under low SNR conditions.

\subsubsection{Impact of Cross-Attention Heads on Generation Performance}
Figure \ref{cdf_multihead_ssim_final} illustrates the impact of different cross-attention head configurations ($H=2, 4, 8$) on the fingerprint generation quality (represented by the CDF of SSIM) at an SNR of 20 dB.
A distinct performance stratification can be observed: increasing the number of heads from 2 to 4 results in a substantial performance leap. The orange curve ($H=4$) is significantly shifted to the right compared to the blue curve ($H=2$), indicating a marked improvement in the structural similarity between the generated and ground-truth fingerprints. This suggests that a low number of heads limits the model's capacity to capture complex spatial correlations.
However, as the number of heads is further increased from 4 to 8 (green curve), the performance gain becomes marginal, with the two curves nearly overlapping. This indicates that 4 attention heads provide sufficient feature extraction capacity to characterize the spatial mapping between Alice and Jack. Further increasing the heads yields diminishing returns while introducing additional computational overhead.
Consequently, to balance performance gains with computational complexity, 4 attention heads are selected as the optimal configuration for this study.

\subsubsection{Fingerprint Generation Performance Compared with Comparative schemes}
Figures \ref{SSIM_Cosine} and \ref{PSNR_NMSE} compare the average performance under a 20 dB SNR condition, providing a comprehensive evaluation of the quality of generated fingerprints. 

Figure \ref{SSIM_Cosine} presents the average SSIM and cosine similarity for each method under a 20 dB SNR condition. CADM achieves the best performance on both metrics, with an SSIM of 0.960 and a cosine similarity of 0.996, indicating excellent preservation of spatial structure and directional consistency in the generated CSI fingerprint. Specifically, compared with time-series-based methods such as GRU and LSTM, which do not utilize channel extrapolation, CADM demonstrates significantly stronger generation capability. GRU and LSTM show moderate structural recovery (SSIM: 0.519 and 0.681, respectively), but their lack of spatial context modeling limits their performance in generating fine-grained channel features. In addition, CADM also outperforms other channel extrapolation-based GAI models. CAVAE achieves the SSIM of only 0.377, as the KL regularization in VAEs tends to over-smooth the latent representation and leads to the loss of fine-grained details. In contrast, the iterative refinement of CADM better captures these complex channel structures. CCMDM incorporates context but lacks feature alignment, resulting in a lower SSIM of 0.260. While CA serves as an ablation experiment without model-driven generation, it still achieves a relatively high cosine similarity of 0.682, thanks to its direct spatial proximity. However, its SSIM drops sharply to 0.099, revealing poor structural matching. This contrast highlights the effectiveness of GAI modeling in learning accurate spatial mappings beyond mere signal-level similarity.

Figure \ref{PSNR_NMSE} further evaluates the fingerprint generation performance in terms of PSNR and NMSE in dB. CADM again achieves the best results, with an average PSNR exceeding 31 dB and NMSE lower than -20 dB, demonstrating strong advantages in both signal fidelity and error suppression. Compared with GRU and LSTM, which rely on temporal modeling without spatial extrapolation, CADM obtains better error control and reconstruction quality. Although GRU and LSTM show moderate PSNR (around 13-16 dB), their NMSE values remain relatively high, reflecting limitations in capturing fine spatial structure.
In addition, while CAVAE serves as a generative comparative scheme, it yields inferior performance on both metrics. Compared to CADM, its PSNR and NMSE drop by approximately 20 dB and 18 dB, respectively, a significant deficit primarily stemming from the inherent optimization conflict in VAEs, where the necessity to balance reconstruction against KL divergence hinders the minimization of element-wise errors unlike the diffusion model's focused noise prediction objective. CCMDM, derived from CADM by disabling cross-attention, retains contextual conditioning but fails to capture fine-grained feature dependencies, resulting in a PSNR and NMSE degradation of about 20 dB and 22 dB relative to CADM. Finally, CA operates as a non-parametric baseline without any model training. Although it leverages direct spatial proximity, its absence of learned inference leads to the most significant performance gap, with PSNR decreasing by nearly 26 dB and NMSE increasing by over 25 dB compared to CADM, highlighting the crucial role of spatial reasoning in GAI-based modeling.

% Figure 10 further analyzes performance from the perspectives of PSNR and NMSE. The Cross Attention Diffusion method also achieves the best results on these metrics, with an average PSNR exceeding 31 dB and NMSE lower than -17 dB, indicating significant advantages in both image fidelity and error control. LSTM and GRU, as temporal models, show moderate performance in error control, whereas Mask Diffusion and VAE exhibit limitations in both accuracy and consistency. It is worth noting that although the Jack-based method requires no training, its performance on both metrics is significantly inferior to generative models, further underscoring the necessity of spatial modeling in channel generation.

\begin{figure}[t]
    \centering
    \includegraphics[width=0.4\textwidth,trim=0 0 0 0,clip]{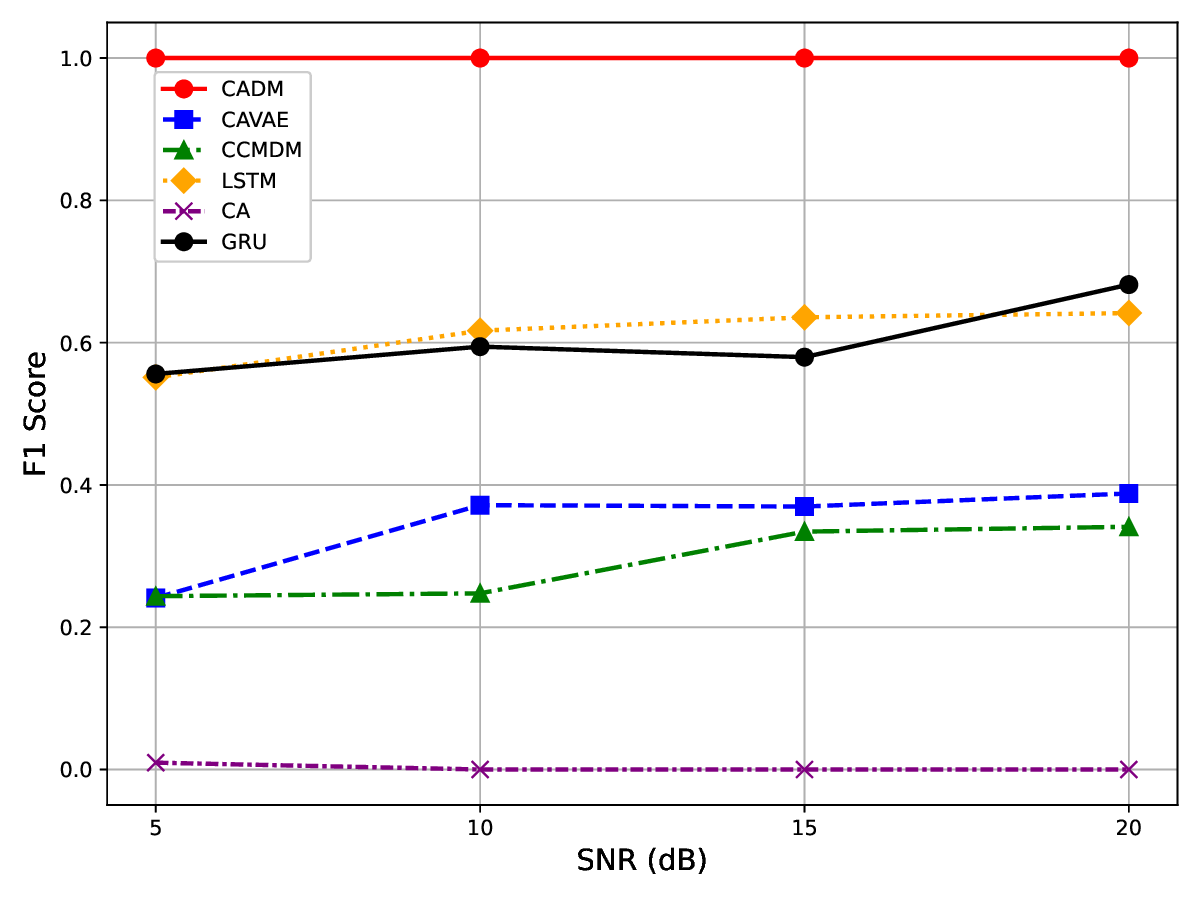}
    \caption{Identity authentication performance of the proposed and comparative schemes at different SNR levels, where Cosine Similarity is used to distinguish between Alice and Eve.}
    \label{COS}
\end{figure}

% \subsection{PLA Performance Compared with Comparative Schemes}

\begin{figure}[t]
    \centering
    \includegraphics[width=0.4\textwidth,trim=0 0 0 0,clip]{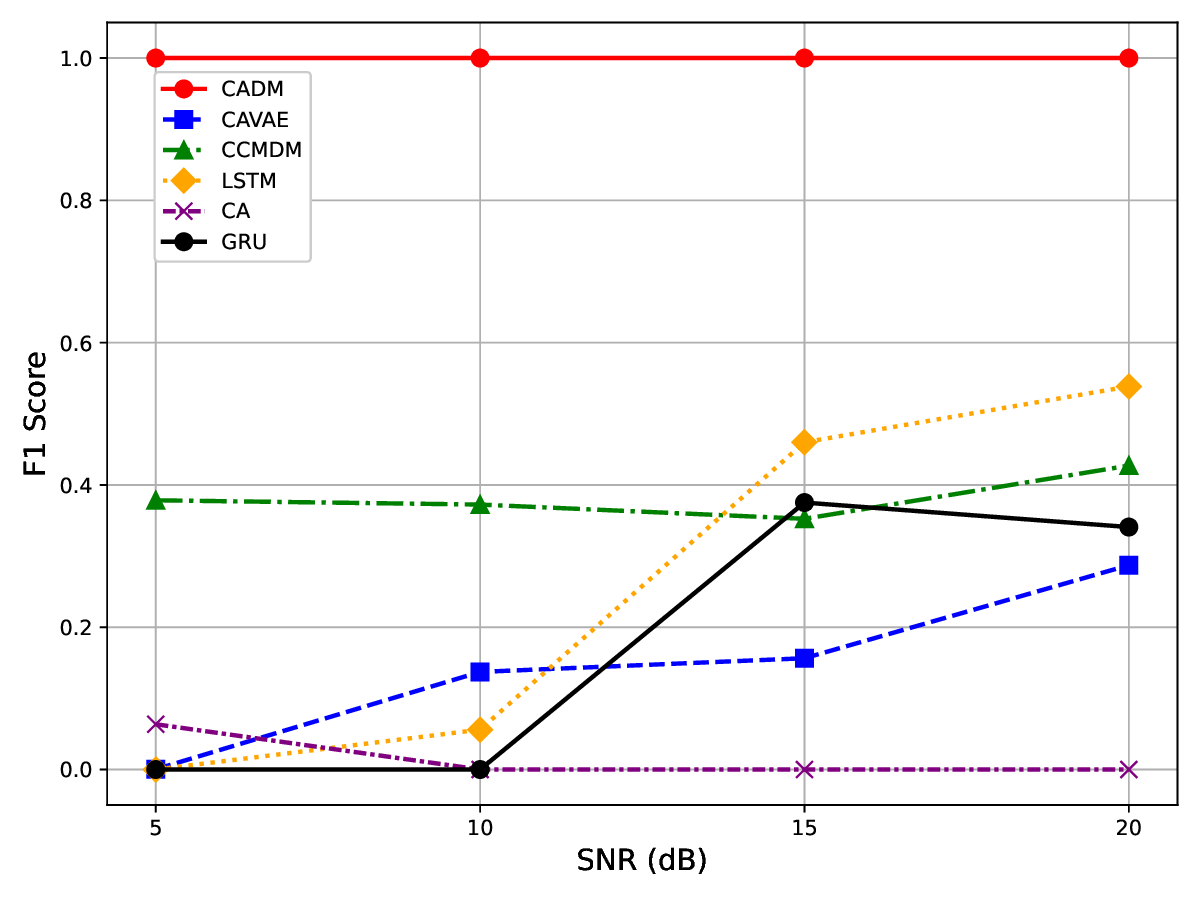}
    \caption{Identity authentication performance of the proposed and comparative schemes at different SNR levels, where PSNR is used to distinguish between Alice and Eve.}
    \label{PSNR}
\end{figure}

\begin{figure}[t]
    \centering
    \includegraphics[width=0.4\textwidth,trim=0 0 0 0,clip]{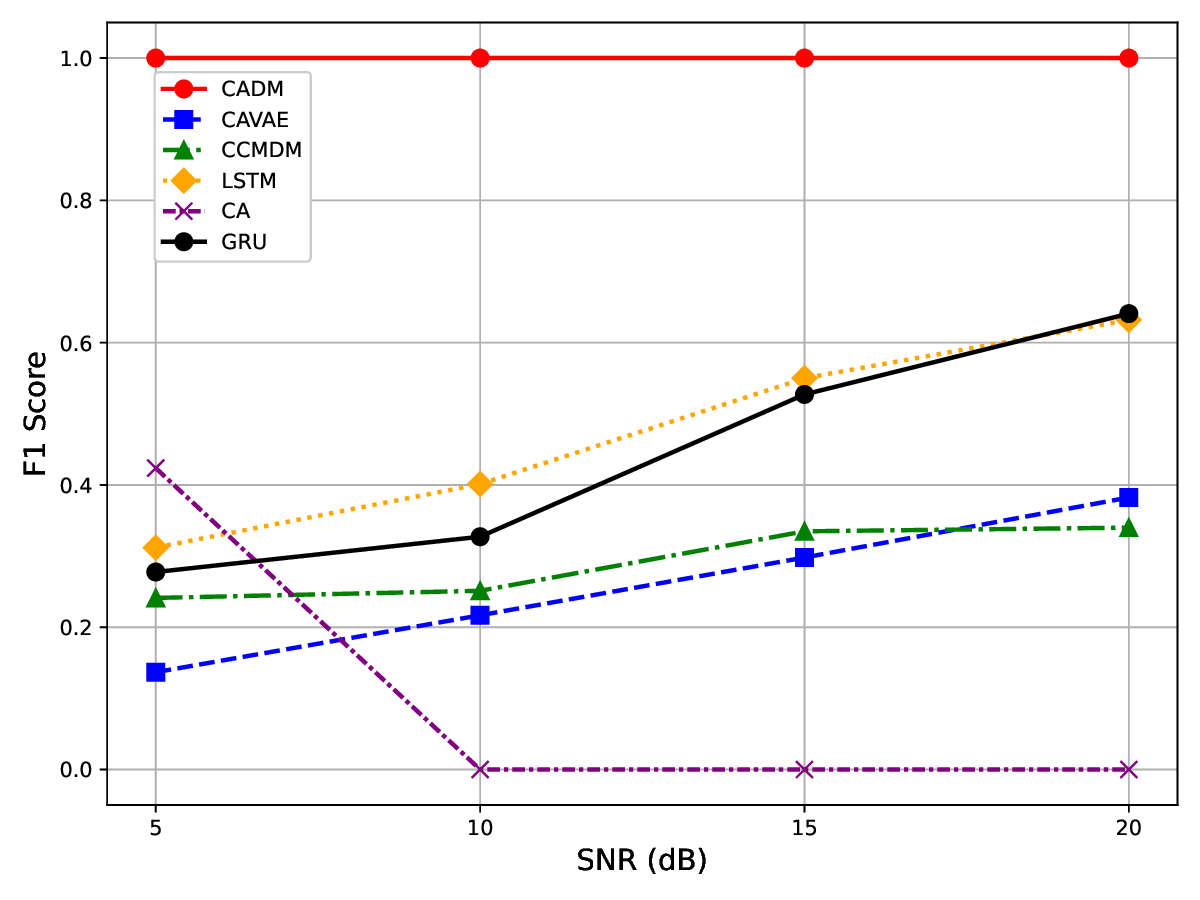}
    \caption{Identity authentication performance of the proposed and comparative schemes at different SNR levels, where SSIM is used to distinguish between Alice and Eve.}
    \label{SSIM}
\end{figure}

\begin{figure}[t]
    \centering
    \includegraphics[width=0.4\textwidth,trim=0 0 0 0,clip]{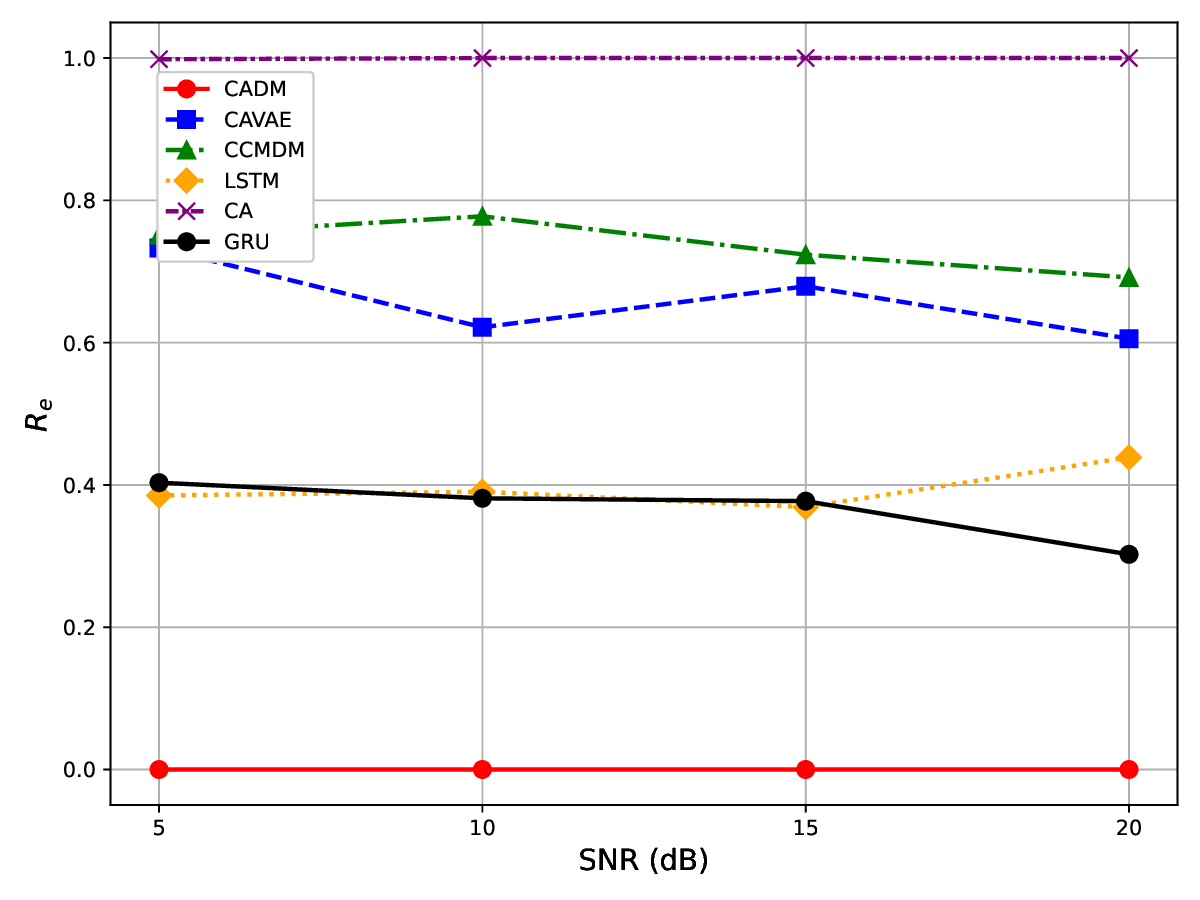}
    \caption{Identity authentication performance of the proposed and comparative schemes at different SNR levels with $R_e$, where Euclidean Distance is used to distinguish between Alice and Eve.}
    \label{Eudis}
\end{figure}

\subsubsection{PLA Performance Compared with Comparative Schemes}

Figure \ref{COS} shows the authentication performance of different schemes evaluated by cosine similarity across various SNR levels. CADM maintains an F1 score of 1.0 consistently, demonstrating strong robustness and effective preservation of directional channel features.
Compared to time-series models like GRU and LSTM, which lack channel extrapolation, CADM achieves significantly better performance. Although GRU and LSTM improve slightly with higher SNR, their overall F1 scores remain lower, indicating the limitations of temporal modeling in capturing spatial structures critical for authentication.
For other models that apply channel extrapolation, such as CAVAE and CCMDM, the performance gap is still clear. These models fail to fully exploit spatial correlations, resulting in reduced accuracy. CA performs the worst, reinforcing the importance of combining extrapolation with GAI modeling to achieve robust authentication under varying noise conditions.
% Figure 11 compares the authentication performance of the proposed method and baseline approaches based on SSIM. In Figure 11, the authentication results based on cosine similarity show that Cross Attention Diffusion consistently achieves an F1 score of 1 across all SNR levels, significantly outperforming other methods. This indicates that the proposed method maintains excellent consistency and stability in preserving the directional structure of the channel. In contrast, LSTM and GRU perform relatively well under medium and high SNRs, but still lag behind overall, while the Jack and Mask Diffusion approaches exhibit significantly weaker authentication capabilities.

% Figure 12 compares the authentication performance based on PSNR. The evaluation results in Figure 12 also show that Cross Attention Diffusion achieves a perfect F1 score, demonstrating excellent amplitude fidelity in the generated channel. LSTM, GRU, and Cross-Att VAE show improving authentication performance at higher SNRs, while Mask Diffusion remains relatively stable and less sensitive to changes in SNR.

Figure \ref{PSNR} illustrates the authentication performance of different schemes based on PSNR under varying SNR conditions. CADM consistently maintains an F1 score of 1.0, showing excellent resilience against noise and strong capability in preserving the fidelity of CSI fingerprints.
Compared with GRU and LSTM, which do not incorporate channel extrapolation, CADM demonstrates a substantial performance advantage. While GRU and LSTM show some improvements at higher SNRs, their F1 scores remain limited, revealing the insufficiency of purely temporal modeling in complex fingerprint inference.
In contrast, CAVAE and CCMDM exhibit noticeable performance gaps. Despite leveraging extrapolation, they fail to deliver consistent authentication accuracy due to weaker representation and feature integration. CA performs the worst across all SNR levels, highlighting the necessity of combining channel extrapolation with a robust GAI model for reliable authentication under diverse channel conditions.

Figure \ref{SSIM} presents the F1 score performance of different schemes evaluated by SSIM across various SNR levels. CADM maintains an F1 score of 1.0 across all SNR conditions, demonstrating outstanding stability and effective preservation of structural similarity in the generated CSI fingerprints.
Compared to GRU and LSTM, which lack spatial extrapolation, CADM delivers a clear performance advantage. Although LSTM and GRU improve gradually as SNR increases, their scores remain notably lower, suggesting that temporal modeling alone is insufficient for capturing the spatial structure required for reliable authentication.
Among the extrapolation-based methods, CAVAE and CCMDM show limited improvements and fail to match the performance of CADM. Despite using extrapolation, these models exhibit lower F1 scores due to their weaker representation learning capabilities. 
CA achieves acceptable authentication performance at 5 dB, due to the identical environmental noise experienced by Alice and Jack which increases the local structural similarity and elevates SSIM. As the SNR increases, the inherent channel differences become more pronounced, lowering the structural similarity and weakening the authentication capability. These results further emphasize the necessity of combining channel extrapolation with powerful GAI models to enhance authentication robustness.

Figure \ref{Eudis} shows the overall error rate ($R_e$) of each scheme versus SNR when Euclidean distance is used for authentication. The proposed CADM holds $R_e = 0$ from 5 dB to 20 dB, showing that it preserves Alice’s CSI fingerprint so faithfully that neither false acceptance nor false rejection arises as noise increases. The time-series comparative schemes, GRU and LSTM, both start near 0.40 at 5 dB. As SNR improves, GRU declines to roughly 0.30, whereas LSTM climbs back to about 0.43 at 20 dB because, once channel noise is suppressed, Euclidean distance becomes dominated by absolute-amplitude mismatches that LSTM cannot model, creating extra errors. The channel extrapolation-based comparative schemes, CAVAE and CCMDM, exhibit an overall downward trend but show noticeable performance fluctuations between 10 dB and 15 dB due to limited generative fidelity and imperfect feature alignment. CA maintains an error rate close to 1.0 across the entire SNR range, confirming that channel-extrapolation relationships are highly implicit and cannot be easily captured without dedicated modeling. Taken together, these observations reveal that only CADM’s precise channel extrapolation coupled with a powerful GAI model drives $R_e$ to zero and ensures robust CSI fingerprint-based authentication.

% \section{Conclusions}

% We have proposed APEG to enable dynamic device identification in future 6G. In this scheme, a collaborator is deployed near Alice, maintaining a fixed relative position throughout Alice’s mobility to support channel extrapolation. Additionally, We have proposed CCMDM for CSI fingerprint generation. Furthermore, we have proposed CADM to align auxiliary features, facilitating accurate user-domain channel extrapolation from the collaborator to Alice. Simulation results have demonstrated the proposed scheme’s superiority than existing schemes. \textcolor{blue}{Consequently, due to its ability to leverage collaborative sensing for robust authentication, the proposed framework shows great potential for deployment in dynamic 6G scenarios, such as Vehicle-to-Everything (V2X) platooning systems.}
% In future work, we will investigate higher-level prior knowledge by extracting environmental factors that influence Alice’s channel characteristics to enhance authentication performance. \textcolor{blue}{Moreover, we plan to relax the strict fixed-position constraint by extending the framework to the Angle-Delay Domain. This extension will allow the collaborator to support channel extrapolation by simply sharing local scattering environments without requiring precise geometric alignment, thereby broadening the applicability of APEG to NLoS and complex dynamic scenarios.}

\section{Conclusion and Future Work}
\label{sec:conclusion}

\subsection{Conclusion}
We have proposed APEG to enable dynamic device identification in future 6G. In this scheme, a collaborator is deployed near Alice, maintaining a fixed relative position throughout Alice’s mobility to support channel extrapolation. Additionally, we have proposed CCMDM for CSI fingerprint generation. Furthermore, we have proposed CADM to align auxiliary features, facilitating accurate user-domain channel extrapolation from the collaborator to Alice. Simulation results have demonstrated the proposed scheme’s superiority than existing schemes. Consequently, due to its ability to leverage collaborative sensing for robust authentication in time-varying channels, the proposed framework shows great potential for deployment in dynamic 6G IIoT scenarios, such as smart factories where fixed sensors must operate in environments with moving scatterers like mobile robots and automated machinery.

\subsection{Future Work}
In future work, we will investigate higher-level prior knowledge by extracting environmental factors that influence Alice’s channel characteristics to enhance authentication performance.While the current CADM generalizes well to new devices and environmental drifts via spatial rule learning and online fine-tuning, we aim to extend the framework to support concurrent multi-device authentication. Moreover, we plan to relax the strict fixed-position constraint by extending the framework to the Angle-Delay Domain. This extension will allow the collaborator to support channel extrapolation by simply sharing local scattering environments without requiring precise geometric alignment, thereby broadening the applicability of APEG to NLoS and complex dynamic scenarios. Furthermore, we intend to further validate and optimize the proposed scheme in high-speed mobility scenarios, investigating mechanisms to mitigate the impact of severe Doppler shifts and rapid time-varying fading to ensure robust authentication. Finally, addressing the security implications of a potentially compromised collaborator is paramount. We will explore trust evaluation mechanisms and multi-collaborator consensus protocols to detect and mitigate malicious behaviors from the helper node, thereby fortifying the APEG framework against internal vulnerabilities.

\bibliography{ref.bib}
\bibliographystyle{IEEEtran}

\vfill

\end{document}